\documentclass[10pt]{article}
\usepackage[utf8]{inputenc}
\pdfoutput=1
\usepackage{natbib}
\usepackage[english]{babel}
\usepackage{hyperref,graphicx,xspace,float,afterpage,longtable}
\usepackage[pdftex]{color}
\usepackage[labelsep=period,labelfont=bf]{caption}
\usepackage{geometry}
\usepackage{latexsym,amsmath,amsfonts,amssymb,textcomp,calc,booktabs,url}

\begin{document}
\thispagestyle{empty}
\begin{center}
\Large Thomas Ruedas\textsuperscript{1,2}\\[2ex]
Doris Breuer\textsuperscript{2}\\[5ex]
\textbf{On the relative importance of thermal and chemical buoyancy in regular and impact-induced melting in a Mars-like planet}\\[5ex]
final version\\[5ex]
9 July 2017\\[10ex]
published in:\\
\textit{Journal of Geophysical Research -- Planets} 122(7), pp.~1554--1579 (2017)\\[15ex]
\normalsize
\textsuperscript{1}Institute of Planetology, Westf\"alische Wilhelms-Universit\"at, M\"unster, Germany\\[5ex]
\textsuperscript{2}Institute of Planetary Research, German Aerospace Center (DLR), Berlin, Germany
\rule{0pt}{12pt}
\end{center}
\vfill
\footnotesize An edited version of this paper was published by AGU. Copyright (2017) American Geophysical Union. That version of record is available at \url{http://dx.doi.org/10.1002/2016JE005221}.
\normalsize
\newpage
\title{On the relative importance of thermal and chemical buoyancy in regular and impact-induced melting in a Mars-like planet}

\author{Thomas Ruedas\thanks{Corresponding author: T. Ruedas, Institute of Planetology, Westf\"alische Wilhelms-Universit\"at, M\"unster, Germany (t.ruedas@uni-muenster.de)}\\{\footnotesize Institute of Planetology, Westf\"alische Wilhelms-Universit\"at M\"unster, Germany}\\{\footnotesize Institute of Planetary Research, German Aerospace Center (DLR), Berlin, Germany}\\[2ex]
Doris Breuer\\{\footnotesize Institute of Planetary Research, German Aerospace Center (DLR), Berlin, Germany}}
\date{}
\maketitle
\textbf{Key points}
\begin{itemize}
\item Compositional buoyancy stabilizes melting-induced anomalies at the base of the lithosphere.
\item The anomalies may be detectable by gravimetry but not by seismics with a sparse station network.
\item Neglecting impact-induced mantle anomalies results in overestimation of the local crustal thickness.
\end{itemize}

\begin{abstract}
We ran several series of two-dimensional numerical mantle convection simulations representing in idealized form the thermochemical evolution of a Mars-like planet. In order to study the importance of compositional buoyancy of melting mantle, the models were set up in pairs of one including all thermal and compositional contributions to buoyancy and one accounting only for the thermal contributions. In several of the model pairs, single large impacts were introduced as causes of additional strong local anomalies, and their evolution in the framework of the convecting mantle was tracked. The models confirm that the additional buoyancy provided by the depletion of the mantle by regular melting can establish a global stable stratification of the convecting mantle and throttle crust production. Furthermore, the compositional buoyancy is essential in the stabilization and preservation of local compositional anomalies directly beneath the lithosphere and offers a possible explanation for the existence of distinct, long-lived reservoirs in the martian mantle. The detection of such anomalies by geophysical means is probably difficult, however; they are expected to be detected by gravimetry rather than by seismic or heat flow measurements. The results further suggest that the crustal thickness can be locally overestimated by up to $\sim 20$\,km if impact-induced density anomalies in the mantle are neglected.
\end{abstract}

\section{Introduction}
In spite of the obvious and well-known geological evidence for their extent and ubiquity, the connection between meteorite impacts and the convective processes that shape the thermal and chemical evolution of planetary interiors still tends to be underappreciated. Among the first numerical studies of mantle convection that considered the effects of large impacts are those by \citet{Rees:etal02,Rees:etal04}, which investigate how a basin-forming impact could introduce a strong thermal anomaly and trigger regional, potentially long-lived convection in a mantle that would otherwise convect only sluggishly. These authors already pointed to the potential of such an impact to seed a long-lived, stable superplume or other sort of thermal anomaly that feeds a major volcanic center, in their case the Tharsis province on Mars, as proposed earlier on the basis of geological evidence \citep[e.g.,][]{PHScGl79}. They also observed that impacts entail strong transient increases in melt production that result in the formation of thickened crust in the area affected by the impact and proposed that this may explain the areoid and topography of Tharsis.\par
\citet{Rees:etal02,Rees:etal04} started their simulations with the impact into a mantle in a simple, symmetric state assumed to represent pre-impact conditions and described the heating and geometry of the impact as summarized by \citet{Melosh89}. In a more recent study, \citet{WAWatt:etal09} have refined the physical model of the impact, in particular with regard to the question how the depth-dependent properties of the target body affect shock propagation and, as a consequence, heating. They also account for the fact that the isobaric core of the impact lies at a certain depth below the pre-impact surface. Their models demonstrate the dynamics of impact-induced plumes: while small events are obliterated quickly in the general flow of the mantle, large impacts cause a strong localized upwelling that results in a large anomaly that spreads beneath the lid and can sweep away cold downwellings. The upwelling induces a flow field that may attract plumes from its neighborhood and merges them, but on the other hand, the anomaly may reduce the convective vigor in the mantle as a whole. \citet{JHRoAr-Ha12} have extended this approach to the effects of multiple basin-forming impacts and also studied the effects of impact size and initial condition in some more detail. \citet{Gola:etal11} investigated the effects of an even larger, dichotomy-forming impact, but their focus lies more on core formation and the very earliest stages of martian evolution.\par
Except for \citet{Rees:etal04} and \citet{Gola:etal11}, the previous workers did not consider compositional effects of impact-induced melting, and the paper by \citet{Rees:etal04} applies very simplistic models for both the impact, which is part of their initial condition, and the dependence of depletion and melt content on melting. On the other hand, various numerical convection studies of martian mantle evolution did include compositional effects to some extent \citep[e.g.,][]{BScho:etal01,Kiefer03,QLiKi07a,KeTa09,OgYa11,OgYa12,Rued:etal13b,Rued:etal13a,PlBr14,Breu:etal16}, but some of them do so only with regard to heat source concentrations, and none include impacts. In particular, \citet{BScho:etal01}, \citet{OgYa11,OgYa12}, and \citet{PlBr14} observed that the compositional density variations produced by melting may be crucial in establishing and preserving long-lived chemical heterogeneities in the martian mantle; a similar observation was made by \citet{Rolf:etal17}, who studied how basin-forming impacts may have altered lunar evolution. Furthermore, the review on the chemical reservoirs in the martian interior by \citet{Breu:etal16} includes some examples of simple local structures with anomalous density or viscosity and discusses their evolution, but these structures do not belong into any specific geological category.\par
In this study we consider both the thermal effects and the compositional changes from regular melting processes and in particular from melting induced by basin-forming impacts of different magnitudes, whereby the compositional changes are derived from a detailed mineralogical model. We compare the relative importance of the individual contributions of temperature and composition to buoyancy on the short-term as well as on the long-term evolution of a planet, which we assume to share many properties with Mars. In addition to the dynamical aspects, we also try to identify the effect of compositional variations on observables, in particular density and seismic velocity.

\section{Method}
\subsection{Convection and mineral physics models}
Our method builds upon or extends the approach described in detail in \citet{Rued:etal13b,Rued:etal13a}, so we limit ourselves to a brief overview and describe shortly the more important changes and extensions. The convection code is a modified version of the code STAGYY \citep{Tackley96a,Tackley08} and solves the equations of conservation of mass, momentum, and energy in the compressible, anelastic approximation on a two-dimensional spherical annulus grid \citep{HeTa08} with $512\times 128$ points and stress-free, isothermal top and bottom boundaries. Melting and the transport of trace components such as water and radionuclides are modeled with tracer particles, with 50 tracers per cell. In impact sites as well as in any other partially molten region, we extract all melt down to the extraction threshold $\varphi_\mathrm{r}$ of 0.7\% if the melt is less dense than the matrix and if there is a contiguous pathway of partially molten rock up to the base of the lithosphere. In terms of the redistribution of incompatible components, the use of such a threshold corresponds to the continuous or dynamic melting discussed in detail by \citet{DMShaw00}. As in our previous work, all extracted melt is instantaneously added to the top to build the crust. Coupled to the convection model is a detailed model of the petrological properties of the mantle consisting of a parameterization of experimental phase diagrams of the martian mantle, its melting behavior, and its thermoelastic properties such as the density, from which the buoyancy forces are derived. The thermoelastic properties are derived from the mineralogical model starting from mineral endmembers and assuming ideal mixing; we replaced the polynomial-based parameterization of these properties from \citet{Rued:etal13a} with a formalism based on Mie--Gr\"uneisen--Debye theory and used a database compiled from literature data from numerous experimental studies; a detailed description of several aspects of the mineralogical, rheological, and chemical model is given in the Supporting Information. Moreover, we have now also implemented a petrological model of basalt/eclogite and its properties at a similar level of detail as the peridotite. The calculation of thermal conductivities takes advantage of the new database by \citet{HoBr15}, which covers many of the relevant minerals, and applies effective medium theory to determine the bulk conductivity; compared with the former dataset, it results in somewhat lower conductivities especially in the deeper parts of the mantle. The mineralogical model is also used for calculating melting partitioning coefficients, whereby we attempted to improve the previous work of \citet{Rued:etal13b,Rued:etal13a} by including dependence on pressure, temperature, and composition where applicable and possible. In particular, a better parameterization of iron partitioning is implemented for several mineral phases, which is an important factor in the relation between depletion and compositional buoyancy. Moreover, the lattice strain model is applied to radionuclide partitioning in clinopyroxenes and garnets \citep[e.g.,][]{BJWoBl14}; a brief summary is given in the Supporting Information. The viscosity of the mantle depends on pressure, temperature, composition, and melt fraction as in the previous work (see Supporting Information). The density of the melt, which needs to be evaluated for several purposes, especially in order to determine whether it is extractable, is now calculated using the new deformable hard-sphere model from \citet{JiKa11,JiKa12} rather than the less versatile model from \citet{ZJiKa09} used in our previous work. As our petrological and melting model is not yet complete enough to yield the complete composition of the melt, we use the martian crust major oxide composition from \citet[Tab.~6.4]{SRTaMcLe09} as input for the melt density model. Furthermore, we have revised our solidus parameterization using new data (see Appendix~\ref{app:sol}).\par
The core enters the model as a thermal boundary condition on the interior boundary that represents the core--mantle boundary (CMB) and is described with an energy balance model following \citet{Nimm:etal04} and \citet{JPWiNi04} but with a more elaborate one-dimensional thermodynamical model of its properties that includes the calculation of adiabatic profiles of temperature and various physical properties for the core; the model is largely identical to that used in our previous work, with the difference that we now use material properties directly derived from experimental data not only for liquid iron, but also for liquid Fe--S alloys \citep{KaTo79,Sanl:etal00,Balo:etal01,Balo:etal03,Nish:etal11}.
\subsection{Impact shock heating and melt production}
The impact process is not included via coupling or interfacing the convection algorithm with a fully dynamical hydrocode simulation but is described using a combination of simpler analytical or semi-empirical approximations. With regard to the energy input from the impact, we follow the approach pioneered by \citet{WAWatt:etal09} in many respects. When considering a large impact event such as the one that created Utopia, we start from the known diameter $D_\mathrm{f}$ of the final crater and use scaling laws for complex craters to work back to determine the diameter of the transient crater that is formed as an immediate consequence of the impact,
\begin{equation}
D_\mathrm{tr}=\left(0.8547 D_\mathrm{f}D_\mathrm{s2c}^{0.13}\right)^\frac{1}{1.13}=0.87028 D_\mathrm{f}^{0.885} D_\mathrm{s2c}^{0.115}\label{eq:Df-tr}
\end{equation}
\citep{Melosh11}, and of the impactor,
\begin{equation}
D_\mathrm{imp}=0.83\left(\frac{\varrho}{\varrho_\mathrm{imp}}\right)^{0.43}D_\mathrm{tr}^{1.28} v_{z,\mathrm{imp}}^{-0.56} g^{0.28}\label{eq:Dimp-tr}
\end{equation}
\citep[e.g.,][]{WeIv15}; the constant pre-factor for the transient crater diameter accounts for the fact that the measured final (complex) crater diameter, which results from the collapse of the transient cavity, is measured from rim to rim rather than at the level of the reference surface and reduces it correspondingly \citep{Melosh11}. Here, $v_{z,\mathrm{imp}}=v_\mathrm{imp}\sin\phi$ is the vertical component of the impactor velocity, $D_\mathrm{s2c}$ is the diameter of the transition between simple and complex craters, which is $\sim$5.6\,km \citep[global mean from][Tab.~3]{RoHy12b} for Mars, $g$ is the acceleration of gravity, and $\varrho$ and $\varrho_\mathrm{imp}$ are the densities of the target and the impactor, respectively; a list of the important symbols is given at the end of this paper. The impactor is assumed to be a rocky (S-type) asteroidal object hitting Mars with an absolute velocity of 9.6\,km/s \citep{Ivanov01}. In the absence of more specific information, the angle $\phi$ is always set to 45\textdegree, which is the most probable impact angle for an isotropic impactor flux \citep[e.g.,][]{GaWe78}. In the use of the scaling laws, we do not distinguish between complex craters and multi-ring basins, because it seems that the assumption of proportional growth is valid for the latter as well \citep{Spudis93}, and numerical models indicate that many characteristics derived from scaling laws apply to large basins without modification \citep{Pott:etal15}.\par
\begin{figure}[ht!]
\centerline{\includegraphics[width=\textwidth]{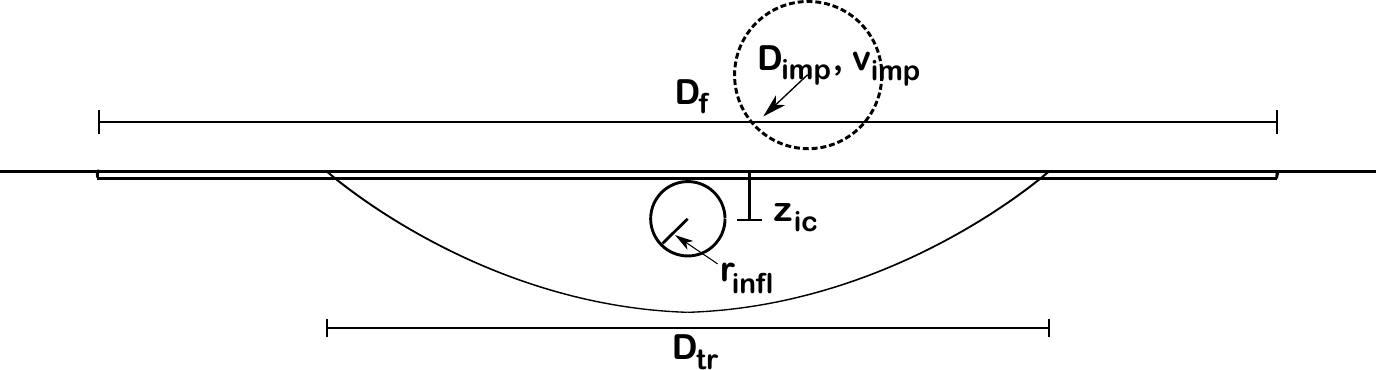}}
\caption{Key variables of the impact model. Final and transient crater depths are not to scale.\label{fig:crater}}
\end{figure}
The most extreme conditions prevail in the central part of the impact region, which has been called ``isobaric core'' by some authors or ``near field'' by others and which is characterized by the near-constancy of the pressure to which the shocked material is subjected. Many previous workers used empirical equations by \citet{Pier:etal97} to determine the radius and depth of this region. We follow them in using their formula for the depth of its center below the surface (cf. Figure~\ref{fig:crater}),
\begin{equation}
z_\mathrm{ic}=0.1525D_\mathrm{imp} v_{z,\mathrm{imp}}^{0.361},\label{eq:zic}
\end{equation}
but apply one of the new parameterizations of the recent low-$v_{\mathrm{imp}}$ data from \citet{MoAr-Ha16} by \citet{Ruedas17} to describe the dependence of shock pressure on distance, because the decay law from \citet{Pier:etal97} becomes increasingly erroneous at the low $v_{z,\mathrm{imp}}$ needed here. These parameterizations have the advantage of describing $p_\mathrm{s}(r)$ accurately with a smooth function for all $r$ and without imposing the impedance-match solution at $r=0$. For an impact of a dunite meteorite onto a dunite target, the ``inverse-$r$'' parameterization by \citet{Ruedas17} can be written in the form
\begin{equation}
p_\mathrm{s}(r)=\frac{a}{b+\left(\frac{2r}{D_\mathrm{imp}}\right)^n}
\end{equation}
with
\addtocounter{equation}{-1}\begin{subequations}\begin{align}
a(v_{z,\mathrm{imp}})&=0.081p_\mathrm{IM} v_{z,\mathrm{imp}}^{1.478},\\
b(v_{z,\mathrm{imp}})&=0.266 v_{z,\mathrm{imp}}^{1.161},\\
n(v_{z,\mathrm{imp}})&=-0.203+1.954\lg v_{z,\mathrm{imp}}
\end{align}
\end{subequations}
and $v_{z,\mathrm{imp}}$ in km/s. We caution that the coefficients for $a$, $b$, and $n$ work reasonably well for the $v_{z,\mathrm{imp}}$ value of approximately 6.8\,km/s assumed here but may not be suited for much higher $v_{z,\mathrm{imp}}$; the reader is referred to \citet{Ruedas17} for a detailed discussion.\par
Previous workers often also used the empirical definition by \citet{Pier:etal97} for the radius of the isobaric core,
\begin{equation}
r_\mathrm{PVM}=0.225D_\mathrm{imp}v_{z,\mathrm{imp}}^{0.211},\label{eq:ricPVM}
\end{equation}
but as the models from which this formula was derived do not seem to cover the remote parts of the shocked regions for low $v_{\mathrm{imp}}$, we suspect that the quality of eq.~\ref{eq:ricPVM} deteriorates in those cases. Therefore, we prefer to define the size of the isobaric core by the position of the inflection point of $p_\mathrm{s}(r)$:
\begin{equation}
r_\mathrm{infl}=\frac{D_\mathrm{imp}}{2}\sqrt[n]{\frac{n-1}{n+1}b}.\label{eq:rinfl}
\end{equation}
The parameter $p_\mathrm{IM}$ is the impedance-match solution, which is used as a scaling factor and is calculated directly from the material properties using the Hugoniot equations as
\begin{equation}
p_\mathrm{IM}=\varrho(v_\mathrm{B}+S_\mathrm{h}u_\mathrm{IM})u_\mathrm{IM},\label{eq:pIM}
\end{equation}
with the density of the target, $\varrho$, in g/cm$^3$, the speed of sound in the target, $v_\mathrm{B}$, in km/s, and the Hugoniot slope $S_\mathrm{h}=(1+\gamma)/2$ being related to the Gr\"uneisen parameter $\gamma$ \citep{Melosh89}. The particle velocity $u_\mathrm{IM} $ at the interface with the impactor is given by
\begin{equation}
u_\mathrm{IM}=\frac{-B_u+\sqrt{B_u^2-4A_uC_u}}{2A_u}\label{eq:uic}
\end{equation}
with
\addtocounter{equation}{-1}\begin{subequations}\begin{align}
A_u&=\varrho S_\mathrm{h}-\varrho_\mathrm{imp} S_\mathrm{h,imp}\\
B_u&=\varrho v_\mathrm{B}+\varrho_\mathrm{imp}(v_\mathrm{B,imp}+2 S_\mathrm{h,imp} v_{z,\mathrm{imp}})\\
C_u&=-\varrho_\mathrm{imp}v_{z,\mathrm{imp}}(v_\mathrm{B,imp}+S_\mathrm{h,imp} v_{z,\mathrm{imp}})
\end{align}
\end{subequations}
\citep[e.g.,][]{Melosh11}; the properties used here are those of the unshocked material, and the subscript ``imp'' indicates a property of the impactor. The speed of sound of the target is derived from the thermoelastic properties as given by the mineralogical model for the planet via $v_\mathrm{B}=\sqrt{K_S/\varrho}$, where we use values for $K_S$ and $\varrho$ averaged over the isobaric core volume. The speed of sound in the impactor is simply determined using Birch's law, $v_\mathrm{B,imp}=2.36\varrho_\mathrm{imp}-1.75$ \citep[e.g.,][]{Poirier00}, using an assumed density $\varrho_\mathrm{imp}$ of 2.7\,g/cm\textsuperscript{3} for mostly rocky (S-type) asteroidal impactors \citep[e.g.,][]{Fern:etal15}.\par
The propagation of the shock into the target and subsequent isentropic unloading put a certain amount of energy into the target and heat it. Reformulating an estimate by \citet{GaHe63}, \citet{WAWatt:etal09} write this impact heating in terms of shock pressure and lithostatic pressure $p_\mathrm{l}$ as
\begin{equation}
\Delta T=\frac{1}{c_p}\left[\frac{\Delta p}{2\varrho S_\mathrm{h}}\left(1-\frac{1}{\Phi}\right)-\left(\frac{v_\mathrm{B}}{S_\mathrm{h}}\right)^2 (\Phi-\ln\Phi-1)\right]\label{eq:dTmelt}
\end{equation}
with
\addtocounter{equation}{-1}\begin{subequations}
\begin{align}
\Phi&=-\frac{2 S_\mathrm{h} \Delta p}{v_\mathrm{B}^2 \varrho} \left(1-\sqrt{\frac{4 S_\mathrm{h} \Delta p}{v_\mathrm{B}^2 \varrho}+1}\right)^{-1}\\
\Delta p&=p_\mathrm{s}-p_\mathrm{l},
\end{align}
\end{subequations}
where $c_p$ is the specific heat capacity. $\varrho$, $c_p$, $v_\mathrm{B}$, $S_\mathrm{h}$, and $p_\mathrm{l}$ are derived directly from the convection and petrological model and are always positive; they are therefore functions of position, in particular of depth, which corresponds to the climbing shock model of \citet{WAWatt:etal09}. We note in passing that the form of eq.~\ref{eq:dTmelt} entails some limitations on $p_\mathrm{s}$ and $p_\mathrm{l}$ in order for $\Delta T$ to be finite, positive, and real. Inspection of the expressions indicates that this requires $\Phi>0$ and that this condition is always met for $p_\mathrm{s}>p_\mathrm{l}$. For $p_\mathrm{s}=p_\mathrm{l}$, $\Phi=1$ by de l'Hôpital's rule, and hence $\Delta T=0$. For $p_\mathrm{s}<p_\mathrm{l}$, there are still real solutions as long as $p_\mathrm{l}<p_\mathrm{s}+v_\mathrm{B}^2 \varrho/(4S_\mathrm{h})$, but $\Delta T<0$, and so we set the thermal perturbation to zero as $p_\mathrm{s}$ drops below the lithospheric pressure.\par
In principle, melt production itself could be estimated from scaling laws \citep[e.g.,][]{BjHo87,Pier:etal97,Abra:etal12}, but as this approach gives only bulk melt volumes but no information about the spatial distribution of the melt, it is not usually the approach taken in convection models. We consider directly where eq.~\ref{eq:dTmelt} leads to supersolidus temperatures and then calculate the amount of melt basically in the same manner as in regular melting processes. However, as the temperatures and melting degrees may be much higher in a giant impact, there will usually exist a region which is much more pervasively molten than normal melt-producing mantle, possibly beyond the rheological solid--liquid transition or even the liquidus. As convection in such a melt pond cannot be handled within the framework of a mantle convection code, we assume that the volume of the model shock-heated above the solidus melts according to the melting parameterization up to an imposed limit of 60\% at which we assume that only a residue without any other fusible components would remain; the most extreme conditions are only reached at most in a small part of the shocked volume. The melt in excess of the extraction threshold $\varphi_\mathrm{r}$ is added to the top in the usual way as material that has cooled to surface temperature upon eruption. The source region is finally set to the temperature corresponding to $\varphi_\mathrm{r}$, i.e., just above the solidus, for the aforementioned technical reasons; a similar numerical crutch for handling the high post-impact temperatures has also been used by previous workers \citep[e.g.,][]{WAWatt:etal09,JHRoAr-Ha12}. Physically, this procedure can be justified by a combination of several processes: much of the waste heat corresponding to eq.~\ref{eq:dTmelt} is consumed by melting and vaporization; the post-impact melt pond is expected to cool much more rapidly thanks to its vigorous convection than the duration of a single timestep of the mantle convection algorithm, which usually corresponds to at least several thousand to a few tens of thousands of years \citep[e.g.,][]{ReSo06}; and a considerable fraction of the heat is extracted from the mantle along with the erupted melt and implicitly assumed to be lost to space as a consequence of our constant-$T$ boundary condition.
\subsection{Impact crater and ejecta layer formation}
Prior to or simultaneously with causing these thermal effects at the impact site, the impact also forms a crater, a process that includes a modification of the local topography and the permanent redistribution of some of the material at the impact site. The transient crater created in the course of the impact is rapidly modified through slumping of the walls and similar processes in the case of larger events that form complex craters. Although the fixed geometry of the numerical grid, the relatively coarse resolution of global models and the complexity of ejecta ballistics prevent us from modeling these processes in any detail, we attempt to obtain an order-of-magnitude estimate of their basic features, because they potentially affect locally the thickness and composition of the crust, which in turn influences certain observables such as heat flow or gravity. We start by determining from which part of the transient crater material is ejected by calculating a number of mass parcel trajectories within the transient crater using the semi-analytical Z model with a buried source \citep{Maxwell77,Croft80,JERich:etal07}; material expelled by jetting during the first contact between impactor and target is not considered, as these jets seem to account for only a few percent of the total emitted material \citep{BCJohn:etal14}. The transport of the ejecta outside the crater is approximated very simplistically in terms of ballistics of independent particles as suggested by \citet{Taub:etal78}, whereby the contributions of different particle sizes to the accumulated ejecta is given by empirical size--frequency distributions \citep{OKeAh85,OKeAh87}. The result of these calculations are estimates of the ejecta thickness as a function of distance from the impact site and of the ejected and displaced volumes; specifically, we can assess which part of the ejecta ends up within the boundaries of the future final crater. The collapse of the transient crater into the final crater cannot be modeled in any detail in our framework, and so we simply ignore the details of the final topography and impose that in the end the transient crater has transformed into a final crater with a level surface, in which slumped material from the transient crater walls and ejecta are all mixed into a uniform body, with the tracer particles being redistributed according to the final distribution of that material; the deposition of ejecta is treated as an influx of material with surface temperature. Each of these simple models is cheap enough in terms of computational effort to be included into the convection model as a subroutine. We note in passing that \citet{Rolf:etal17}, who also gave consideration to the role of ejecta, describe ejecta blanket formation more straightforwardly in terms of empirical scaling laws but chose to assume zero thickness for ejecta inside the basin.

\section{Results}
\subsection{Model set-up}
In recent years, models of the martian interior based on the analysis of geodetical and gravity data combined with petrological models have helped to tighten the constraints on the size of the martian core. \citet{KhCo08} compiled some interior models, some of which featured a core size small enough to allow for a basal layer of perovskite (pv) and ferropericlase (fp) in the mantle. However, their own preferred model does not include such a layer, and the more recent investigations by \citet{Kono:etal11} and by \citet{Rivo:etal11} also point to a rather large core that would preclude the existence of a (pv+fp) layer; the latter study finds that such a layer would be marginally possible only in a hot mantle. Moreover, the impacts under consideration are not large enough to penetrate to depths that allow them to affect the core--mantle boundary region directly, so that the immediate dynamical effects would unfold in regions quite remote from a (pv+fp) layer. For these reasons, we decided to focus on models with a large core and no basal layer.\par
\begin{table}
\caption{Model parameters. Default values of variable parameters are printed in italics.\label{tab:modpar}}
\centering
\begin{tabular}{lc}\hline
Planetary radius, $R_\mathrm{P}$&3389.5\,km\\
Total planetary mass, $M$&$6.4185\cdot10^{23}$\,kg\\
Surface temperature&218\,K\\
\textit{Mantle}\\
Mantle thickness, $z_\mathrm{m}$&1659.5\,km\\
Initial potential temperature, $T_\mathrm{pot}$&1700\,K\\
Initial core superheating&150\,K\\
Surface porosity, $\varphi_\mathrm{surf}$&0.2\\
Melt extraction threshold, $\varphi_\mathrm{r}$&0.007\\
Bulk silicate Mars Mg\#&0.75\\
Present-day K content&305\,ppm\\
Present-day Th content&56\,ppb\\
Present-day U content&16\,ppb\\
Initial bulk water content&\textit{36}, 72, 144\,ppm\\
\textit{Core (average material properties)}\\
Core radius&1730\,km\\
Sulfur content&16\,wt.\%\\
Thermal expansivity\textsuperscript{a}, $\alpha_\mathrm{c}$&3.5--$4.3\cdot10^{-5}$\,1/K\\
Isobaric specific heat, $c_{p\mathrm{c}}$&750 J/(kg\,K)\\
Thermal conductivity\textsuperscript{a}, $k_\mathrm{c}$&23.5--25.1\,W/(m\,K)\\
\hline
\multicolumn{2}{l}{\textsuperscript{a}Varies with time.}
\end{tabular}
\end{table}
The chemical model assumed for the bulk mantle and crust is the one by \citet{WaDr94}, although we do allow for some variability in the water content, having in mind that the water content of the martian mantle is an unsettled issue and may be on the order of several hundred parts per million \citep[e.g.,][]{McCu:etal10,McCu:etal12a}, i.e., much higher than the 36\,ppm suggested by \citet{WaDr94}. We assume this latter value in the major set of models, i.e., a fairly dry initial bulk silicate water content. In addition to these models, whose mantle quickly becomes almost dry, we also consider two further model series with initial bulk water contents of 72 and 144\,ppm, i.e., two and four times the default value, respectively. This is enough to leave water contents at the tens-of-ppm level in the mantle, even though it lies lower than the values proposed for instance by \citet{McCu:etal12a}; it may therefore correspond more closely to the scenarios suggested by \citet{WaDr94} or more recently by \citet{McCu:etal16a} for the shergottite sources. At these concentrations, the effect of water on thermoelastic properties is not considered to be significant, and even the effect on the position of the solidus will be minor (although we do include it), but it is expected to lower the viscosity of the material noticeably, which is our main motivation for giving it consideration. For the core, we assume a sulfur content of 16\,wt.\%, more similar to those favored by \citet{Rivo:etal11}, i.e., slightly higher than \citet{WaDr94}.\par
We assume an initial state prevailing at 4.4\,Ga that is defined by mantle that is compositionally homogeneous and well-mixed on large scales but is also already depleted by the production of the initial primordial crust; this state is implemented as a random pattern of low-degree depletion ($\sim 0.01$ on average) throughout the mantle that signifies that the entire mantle has experienced melting or has been mixed with residue from melting thoroughly in the transition from the magma ocean. The thickness of the resulting initial crust is not well constrained by observations but it is mostly dependent on the initial temperature.\par
As the objective of this study is the comparison of the contributions of thermally and compositionally induced density variations to the buoyancy, the models come in pairs that include one model each in which the compositional contribution to the buoyancy is suppressed by assigning the relevant physical properties of the unmolten material to the depleted material. In other words, in the ``thermal-only'' (T) model of each model pair, it is assumed that melting does not change the mineralogy and the iron content and hence also leaves density, thermal expansivity, etc. of the residue unchanged, whereas in the other model of the pair (TC), all these properties are functions of the melting degree $f$. However, we do treat the melting material in the T models as depleted residue in the usual sense as far as trace component partitioning or rheology are concerned, i.e., radionuclides and water are removed from the residue in all models.\par
The reference setup is a Mars-like planet with an initial potential mantle temperature $T_\mathrm{pot}$ of 1700\,K, which is in the range deemed reasonable, e.g., with regard to the resulting crustal thickness, by previous studies \citep[e.g.,][]{HaPh02,Mors:etal11}. In most of the models with an impact, the reference model is struck by an S-type asteroid with a velocity of 9.6\,km/s at an angle of 45\textdegree\ at 4\,Ga, which corresponds to 400\,My model evolution. The age of the impact is reasonably close (considering uncertainty) to the ages of three martian craters of different size whose parameters we used to define the model impacts, namely, Utopia, Isidis, and Huygens (see Table~\ref{tab:impacts}); these three impacts are not intended to be faithful representations of the eponymous craters but serve as representatives covering a range of magnitudes that reaches from having a marginal effect on the mantle to the largest observed basin-forming event. Moreover, the planet has already evolved to some extent at this time and has formed a thermal lithosphere of a certain thickness; the most vigorous stage of regular melt production has already come to an end, so that the impact effects will stand out more clearly. Given crustal thicknesses around 79\,km at the time of impact and a melt production zone for regular melting reaching down to $\gtrsim 330$\,km but reaching the strongest depletion at depths $\lesssim 200$\,km, these three impacts have the centers of their isobaric core toward the bottom of the main melting zone (Utopia), near its top (Isidis), and within the crust (Huygens). The isobaric cores correspondingly extend into the mantle below the melting zone in the case of Utopia, cover the crust and melting zone in the case of Isidis, and are limited to the crust in the case of Huygens, which should permit to reveal a possible effect of the relative positions of the pre-existing regular global melting zone and the impact-generated anomaly. An effort was made to place the site of every impact so that it is located neither directly above a hot upwelling nor above major cold downwellings; as demonstrated by \citet{JHRoAr-Ha12}, the pre-existing convective features at the impact site can modify the effects of the impact, although its influence for instance on the impact-generated anomaly in the global heat flow does not exceed 10\%. For this reason, the impacts appear at different positions in different models, as the different buoyancy mechanisms and water contents result in different convection patterns. An overview of all models is given in Table~\ref{tab:models}.\par
The timing of an impact is potentially important, especially in very young planets in which the lithospheric cooling rate is still high. In order to probe this dimension of parameter space, we ran two additional model series in which the impact occurred at 50 or 150\,My model time; the uncertainties in crater-statistics dating certainly warrant assuming different times. Because of the small effect of Huygens-sized events, we considered only Isidis- and Utopia-sized impacts in those cases. Apart from the timing, the model parameters are identical to the 400\,My cases. Physically, the difference in the ``earlier'' cases is that the impactor meets a target that is already quite similar to that in the normal Isidis case as far as the extent and depletion of the regular melting zone and the thickness of the crust are concerned, but the thickness of thermal lithosphere is substantially smaller, such that the target volume is hotter on average.
\begin{table}
\caption{Model impacts. Final crater sizes are from the compilation by \citet{JHRobe:etal09} after \citet{Frey08}, ages are from the same source and from \citet{Werner08}. While we define the radius of the isobaric core according to eq.~\ref{eq:rinfl}, we also give the radius according to \citet{Pier:etal97} (eq.~\ref{eq:ricPVM}) for reference.\label{tab:impacts}}
\centering
\begin{tabular}{lcccccc}\hline
Crater&$D_\mathrm{f}$ (km)&$D_\mathrm{imp}$ (km)&$z_\mathrm{ic}$ (km)&$r_\mathrm{PVM}$ (km)&$r_\mathrm{infl}$ (km)&age (Ma)\\\hline
Huygens&467.25&70.6&21.5&23.8&19.5&3980\\
Isidis&1352&243.5&74.1&81.9&67.1&3810--3960\\
Utopia&3380&699.2&212.9&235.2&192.6&3800--4111\\\hline
\end{tabular}
\end{table}
\begin{table}
\caption{Model pairs\label{tab:models}}
\centering
\begin{tabular}{lccc}\hline
Model pair name&Impact size&$C_\mathrm{H_2O}$ (ppm)&time of impact\\\hline
Ref-36&--&36&--\\
H-36&Huygens&36&400\,My\\
I-36&Isidis&36&400\,My\\
I-36m&Isidis&36&150\,My\\
I-36e&Isidis&36&50\,My\\
U-36&Utopia&36&400\,My\\
U-36m&Utopia&36&150\,My\\
U-36e&Utopia&36&50\,My\\
Ref-72&--&72&--\\
I-72&Isidis&72&400\,My\\
I-72m&Isidis&72&150\,My\\
I-72e&Isidis&72&50\,My\\
U-72&Utopia&72&400\,My\\
U-72m&Utopia&72&150\,My\\
U-72e&Utopia&72&50\,My\\
Ref-144&--&144&--\\
I-144&Isidis&144&400\,My\\
%I-144m&Isidis&144&150\,My\\
I-144e&Isidis&144&50\,My\\
U-144&Utopia&144&400\,My\\
U-144m&Utopia&144&150\,My\\
U-144e&Utopia&144&50\,My\\
\hline
\end{tabular}
\end{table}

\subsection{Dynamical evolution}
The reference models start with a regular pattern of upwellings and downwellings that breaks down after a few tens of millions of years and transitions into a rather irregular flow; the impacts of the ``early'' model pairs occur at a time when this transition begins. The Ref-36 models (see Table~\ref{tab:models}) slowly evolve from the chaotic stage toward the final convection pattern with five to seven stable plumes beginning after $\sim 500$\,My, i.e., shortly after the time when the regular impacts would strike, whereas this final pattern begins to emerge several hundreds of millions of years later with increasing water content of the mantle.\par
\begin{figure}[ht!]
\centerline{\includegraphics[width=0.7\textwidth]{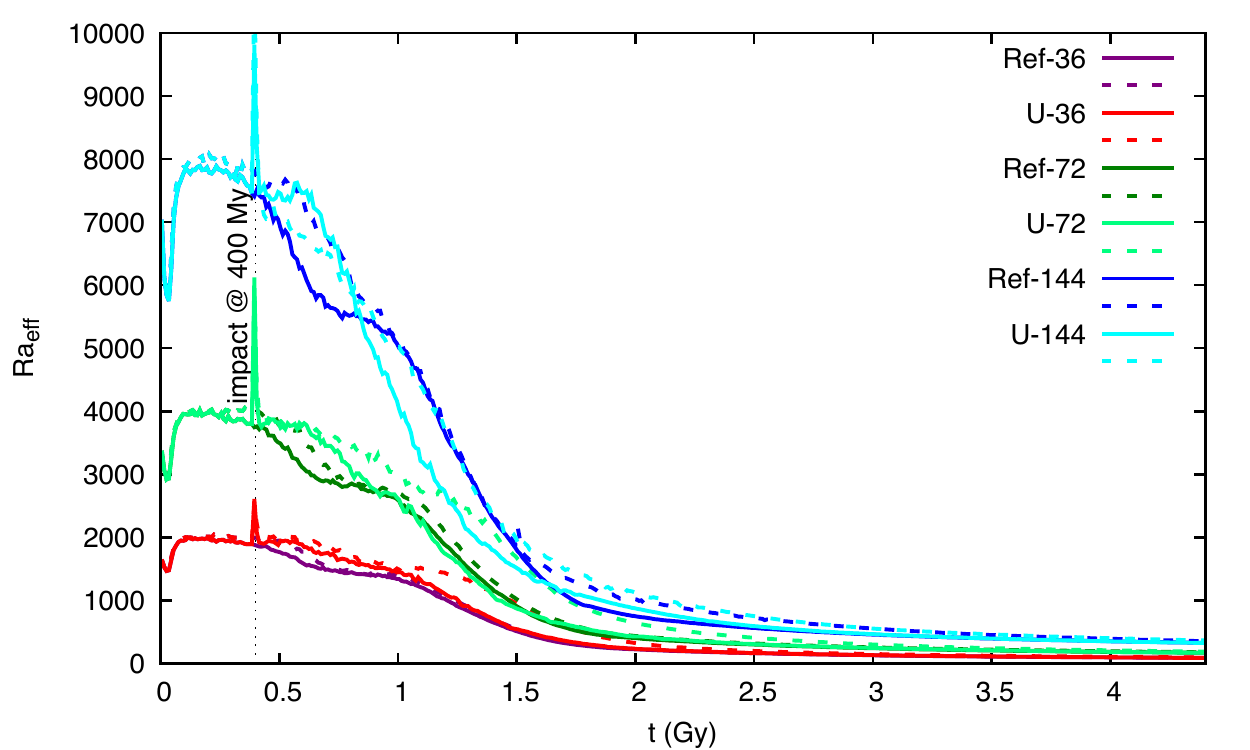}}
\caption{10-My averages of the effective Rayleigh number as a function of time for the reference models and the models with a Utopia-like impact at 400\,My. Models with thermal and compositional buoyancy (TC) are plotted with solid lines and models with purely thermal buoyancy (T) with dashed lines.\label{fig:Raeff}}
\end{figure}
In terms of the (effective) Rayleigh number $Ra_\mathrm{eff}=Ra_0/\eta_\mathrm{mean}$, the evolution of the reference models can be divided into two major phases (Figure~\ref{fig:Raeff}): an early period of vigorous but declining convection that lasts until $\sim 1.5$\,Gy and a later period of nearly constant and calm flow that encompasses the remaining almost 3\,Gy. Irrespective of the water content, the T model variant generally convects a bit more energetically during the early stage, whereby the contrast with the TC variant is more pronounced in the model with low water content (36\,ppm). The Ref-36 models convect at an approximately constant vigor during the first 1\,Gy, whereas the Ref-72 and Ref-144 models show a steep two-stage decline in their Rayleigh numbers, the first of which coincides essentially with the end of their principal period of crust formation. The mean mantle temperatures differ by less than $\sim10$\,K in most cases and decrease almost linearly throughout the evolution. Only the T variant of Ref-144 follows a somewhat different path, because the combination of the generally lower viscosity and the relatively high density of the cooling uppermost mantle let it become gravitationally unstable and develop numerous downwellings, whereas its TC counterpart is stabilized by the additional compositional buoyancy in the melting zone.\par
\begin{figure}[ht!]
\begin{minipage}{\textwidth}
\hspace{0.22\textwidth}TC\hspace{0.49\textwidth}T
\end{minipage}\\[1ex]
\includegraphics[width=\textwidth]{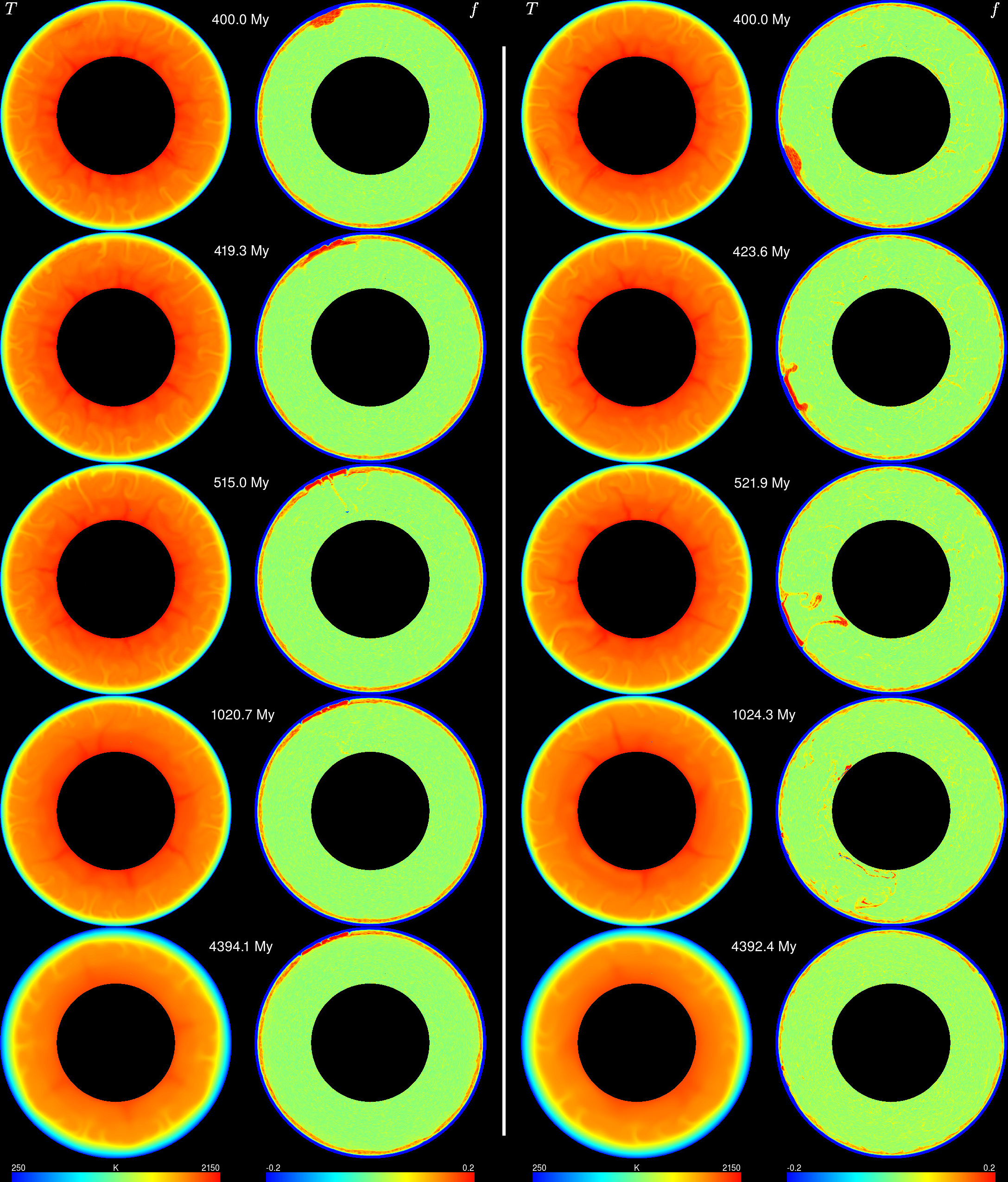}
\caption{Evolution of temperature and composition in the model pair I-36. Left: model with thermal and compositional buoyancy; right: model with purely thermal buoyancy. In the images of composition, a positive melting degree $f$ indicates depleted material; the crust is displayed as strongly ``enriched'' (dark blue). The colorbars are clipped for clarity. Animations of the evolution of both models are available online as Supporting Information.\label{fig:isidis}}
\end{figure}
\begin{figure}[ht!]
\includegraphics[width=\textwidth]{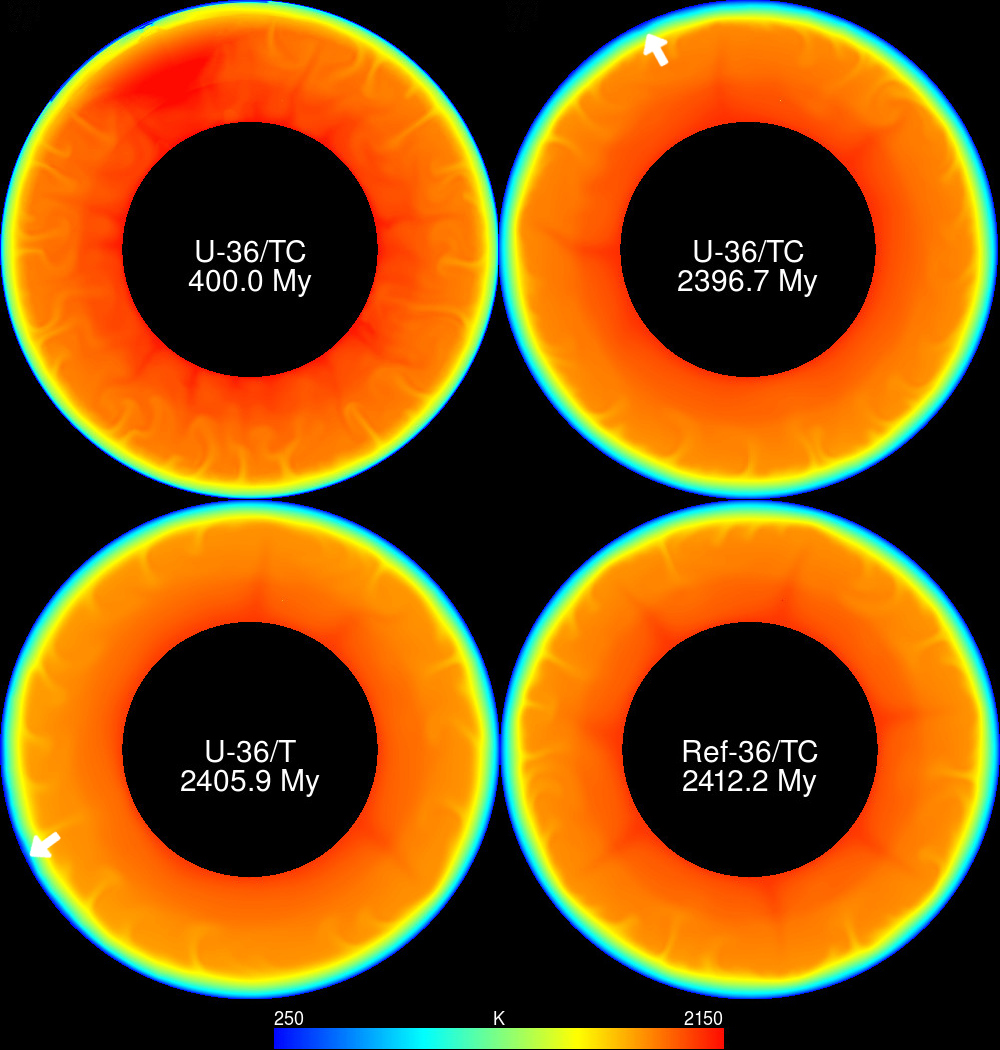}\\
\caption{Temperature field of the TC model from U-36 at the time of the impact, and temperature fields of models U-36 (TC), U-36 (T), and Ref-36 (TC) about 2\,Gy later. The white arrows mark the impact sites. The colorbar is clipped for clarity.\label{fig:plumes2Gy}}
\end{figure}

An impact introduces an instantaneous local thermal disturbance that results in enhanced local heating and melting. This disturbance is visible as a spike signal in several parameters that characterize the thermal evolution, e.g., in the curves of $Ra_\mathrm{eff}$ in Figure~\ref{fig:Raeff}, whereby the magnitude of the signal grows with increasing water content/decreasing viscosity, because a low-viscosity system reacts more strongly to any perturbation. In the models with an earlier impact, the evolution of $Ra_\mathrm{eff}$ is essentially the same, although in the 50\,My series the spike of the impact is absorbed into the general initial steep increase of this parameter. As a consequence of the impact, a compositional anomaly characterized by enhanced depletion develops, as can be seen in the composition ($f$) images in Figure~\ref{fig:isidis}. The sequences in that figure and the corresponding animations in the Supporting Information show how the compositionally buoyant impact anomaly essentially stays in place for billions of years in the TC models, whereas it is dispersed within a few hundreds of millions of years in the T models. The density decrease due to heating and, in the TC models, due to the loss of higher-density constituents triggers a localized upwelling whose influence can reach far beyond the volume immediately affected by the impact itself. In the larger impacts, a hot pulse rapidly spreads out beneath the lithosphere and reinforces melting and crust formation if the center of the isobaric core lies deep enough. In a very large impact, the upwelling modifies the flow field on a regional scale such that pre-existing plumes may be drawn toward the impact site and merge. In most models with larger impacts, this effect persists for several hundreds of millions of years, i.e., much longer than the actual lifetime of the thermal anomaly of a few tens of millions of years, suggesting that a very large (Utopia-sized) impact can enforce a fairly stable regional flow field that is self-sustaining to some extent under favorable conditions; the TC model from U-36 establishes such a long-lived plume for billions of years, as can be seen in Figure~\ref{fig:plumes2Gy} (top row). In a more vigorous convective regime as in its wetter counterparts from U-72 and U-144, the enhanced stirring disturbs the stability of the structure and shortens its lifetime the more the higher the water content is, although the direct hot-material surge from the impact spreads further due to the lower viscosity. In the model series with earlier impacts, there is a tendency of globally reinforced crust production and delamination, but the effect is insignificant except for the 144\,ppm water model of the Utopia-sized event, in which a substantial amount of crustal material is reintroduced into the mantle. In the corresponding T models, the local effect decays much faster, on timescales of severals tens rather than hundreds of millions of years because of the absence of the self-stabilizing compositional anomaly, although the general reinvigoration of convection tends to be a bit more persistent than in the TC models, judging from Figure~\ref{fig:Raeff}. Figure~\ref{fig:plumes2Gy} (bottom left) shows that there is no plume near the former impact site 2\,Gy after the impact, although the convection pattern is nonetheless not as regular as that of the reference model shown in Figure~\ref{fig:plumes2Gy} (bottom  right). This underscores that the compositional buoyancy of the impact-induced anomaly must contribute significantly to the establishment of the stable flow field. An effect is also well visible in the TC models of Isidis-size impacts, but it is less long-lived and extensive, whereas it is insignificant in the T versions. In the Huygens-size events (H-36), which are essentially confined to the thermal lithosphere, there is no effect beyond a small, short-lived local disturbance.\par
A large part of the crust~-- approximately 59\,km~-- is produced in the initialization step as ``primordial'' crust, and crustal growth is strongest at the beginning and declines toward zero afterwards. In the reference models the crust usually reaches its final thickness after 1--1.5\,Gy, reaching values higher by a few kilometers in models with higher mantle water contents. The amount of crust added after initialization lies between approximately 20 and 25\,km in most cases; only the T variants of the models with 144\,ppm water experience more voluminous and protracted crust formation thanks to the reinforcement of convection by the cold downwellings. In some regions that have been fed by plumes over extended periods, the crust eventually grows thick enough to delaminate; nonetheless, delamination is generally not a quantitatively important process, and the small amount of delaminated material is mixed into the deeper mantle within a few tens of millions of years. In Ref-144/T and its impact-affected derivatives delamination is more extensive, because the downwellings drag not only lithospheric mantle but also a part of the thick crust with them, but it does nonetheless not reach a global scale that affects the entire crust.\par
\begin{figure}
\includegraphics[width=\textwidth]{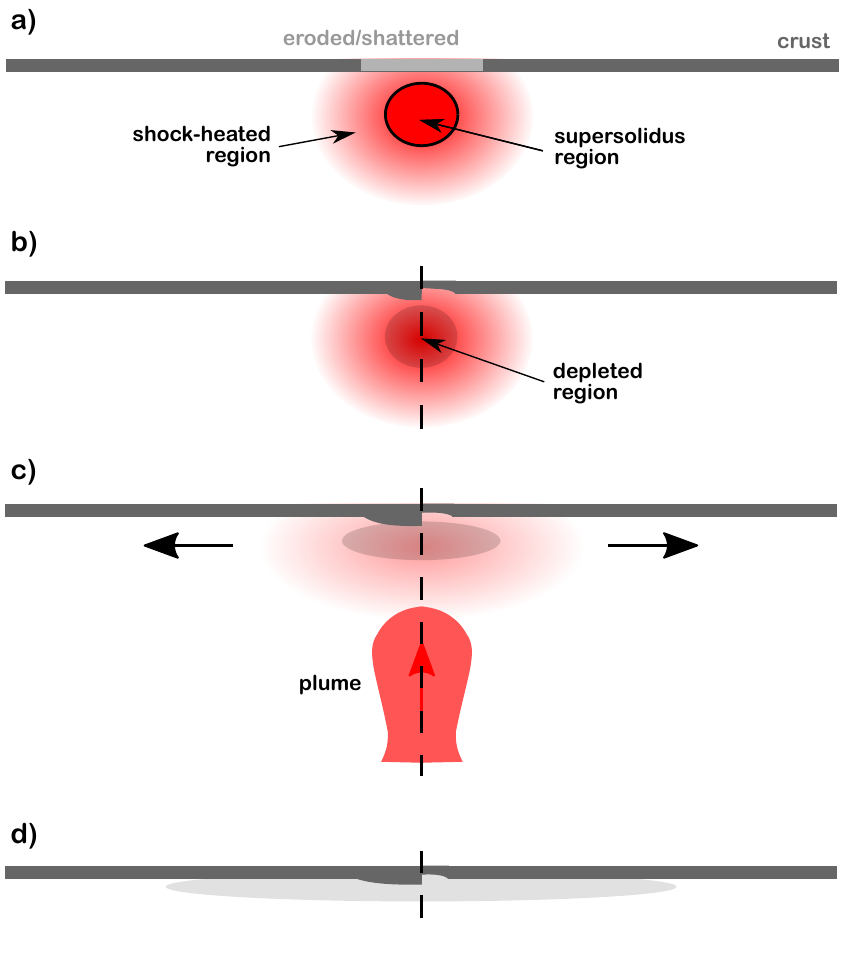}
\caption{Post-impact evolution of a thermochemical anomaly generated by a basin-forming impact. a) Shortly after the impact, the crust around the impact site is removed or shattered, and a large region around the center of the isobaric core is shock-heated, partly to temperatures above the solidus. b) The melt extracted from the supersolidus region leaves behind a depleted residue in the mantle and has formed a new crust together with pre-impact crustal material from the final crater. In the left part of this and the following stages, the new crust is thicker than average, in the right part it is thinner; which variant develops in any given crater depends on the magnitude of the impact and the amount of melt produced by it. c) The thermochemical anomaly rises due to its buoyancy and may trigger a plume from the CMB or attract one from nearby. With time the thermal component of the anomaly fades, whereas the compositional is not much diminished. The rising plume may produce additional crust. d) Final stage: the thermal anomaly and the plume have disappeared, whereas the compositional anomaly has spread out at the base of the lithosphere/crust and remains in place. Variations in the thickness of the crust may be levelled to some extent by creep. The figures are not to scale.\label{fig:anomaly}}
\end{figure}

The enhanced melting and crust formation at an impact site may also result in a thicker crust, but on the other hand, the removal of crustal material by crater formation may compensate to some extent the growth of the crust. In all models, we find that the T variant develops a thicker post-impact crust at the impact site than its TC counterpart, because the buoyant residue from the impact was less stable and hence less efficient at preventing the ascent of fertile material into the melting zone. The post-impact thickening is largely due to the immediate melt production from the impact in the TC models and therefore finishes within a few tens of millions of years, whereas in the T models, the enhanced crustal formation persists for hundreds of millions of years. The extent to which the crust grows depends on the water content of the mantle and the time of the impact: early impacts and/or higher water contents generally entail a thicker crust, because the lower viscosity allows the mantle to convect more vigorously and produce more melt. In the two larger sorts of impact, the post-impact crust is always thicker than the mean of the reference model due to the additional melt production, whereas the Huygens-sized events do not produce enough melt to compensate the removal of crust by the impact itself. Crustal thickness variations exist directly after the impact, but they tend to flatten out soon by spreading laterally as a consequence of the buoyancy forces of the crust and the underlying mantle anomaly, as well as the action of shear forces from divergent mantle flow at its base. Post-impact melt production and crust formation are facilitated by thermal ``rejuvenation'' of the lithosphere caused by the impact, which produces a local pit-like low-viscosity zone with a lifetime of several tens of millions of years in the boundary region between crust and mantle. An example for this phenomenon is visible at the impact sites of the I-36 models in Figure~\ref{fig:isidis}. This process seems to be more efficient in the T models, where the root is less protected by a stable density anomaly in the underlying mantle. Figure~\ref{fig:anomaly} summarizes the principal stages of the post-impact evolution after a large basin-forming impact.

\subsection{Observables}
A variety of geophysical and geochemical observables can be derived from the models. Although it is not our primary goal to reproduce in detail observations from Mars, we will draw comparisons with results from missions to that planet in the following and concentrate on some observables for which data are relatively abundant or which are expected to become available in the near future as results from the InSight mission.
\subsubsection{Surface heat flow, $q_\mathrm{s}$, and elastic thickness of the lithosphere, $z_\mathrm{el}$}
The heat flux estimates from spacecraft missions have so far usually been derived from flexure or admittance models and are mean heat flows through the elastic lithosphere; comparable fluxes from numerical models could be derived using the temperature difference between the top and bottom of the elastic lithosphere, its thickness, and its average thermal conductivity, as done by \citet{Rued:etal13b,Rued:etal13a}. The HP$^3$ experiment on the InSight mission, by contrast, is a real heat flux probe designed to determine the actual surface heat flux at the landing site; this is not identical to the mean lithospheric heat flow, but it can be approximated in a numerical model by a finite-difference formulation using the outermost grid cell layer. In this study, we consider the surface heat flow and the elastic thickness of the lithosphere, which is determined as an intermediate step in the calculation of the mean lithospheric heat flow and should be more readily comparable to independent data as it requires fewer assumptions.\par
\begin{figure}[ht!]
\includegraphics[width=\textwidth]{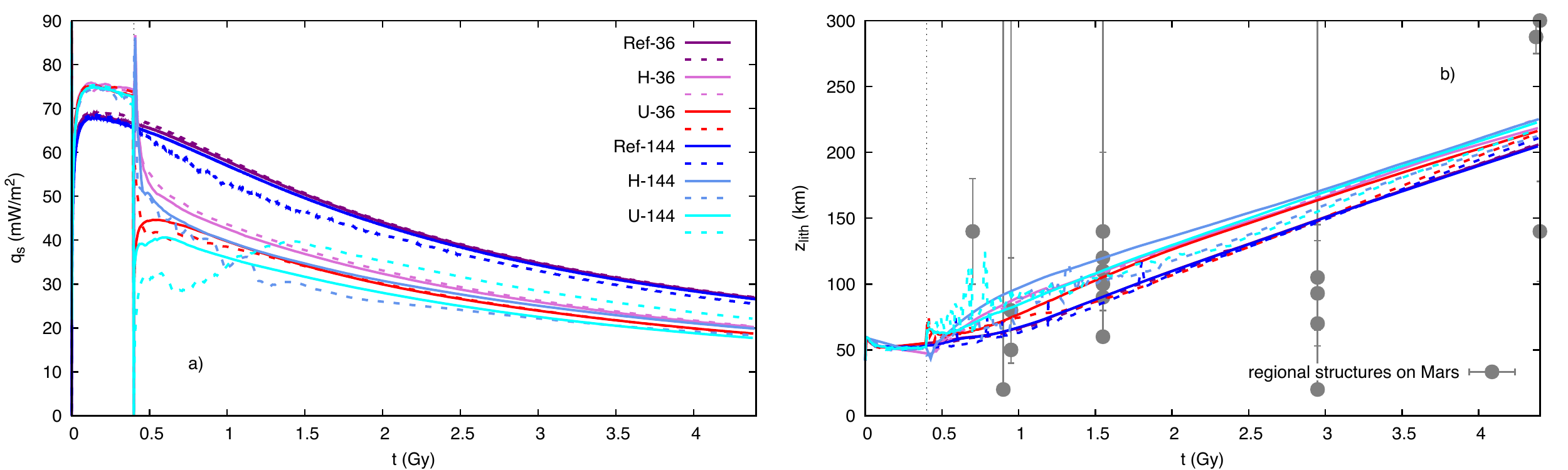}
\caption{Surface heat flow (a) and elastic thickness of the lithosphere (b) of selected models as a function of time; the curves for the Reference models are global means, the curves for the models with impacts show the average within the final crater. TC models are plotted with solid lines, T models with dashed lines. Observation-based elastic thickness values are for regional structures \citep{McGo:etal02,BaNi02,Bell:etal05,RJPhil:etal08,Wieczorek08,RiHa09}. The ages are averages, in most cases based on the age estimates of the corresponding era, and will usually have uncertainties on the order of hundreds of millions of years.\label{fig:qszl_t}}
\end{figure}

The determination of the elastic thickness of the lithosphere from flexure models usually involves the use of non-Newtonian rheology, which is not part of our numerical model. A direct comparison is therefore not possible, but one can still attempt an estimate by assuming an average lithospheric strain rate $\dot\varepsilon$ and a yield stress $\sigma_\mathrm{y}$ typical for lithospheric materials. We follow \citet{GrBr08a}, who assumed a strain rate of $10^{-17}$\,s$^{-1}$ and yield stress of 15\,MPa and then solve the Arrhenius-type flow law for non-Newtonian creep for temperature:
\begin{equation}
T_\mathrm{bd}=\frac{Q}{R_\mathrm{gas}}\left[\ln\left(\frac{B\sigma_\mathrm{y}^\nu}{\dot\varepsilon}\right)\right]^{-1},\label{eq:Tbd}
\end{equation}
where the factor $B$, the activation enthalpy $Q$, and the exponent $\nu$ are $3.1\times10^{-20}$\,Pa$^{-3.05}$s$^{-1}$, 276\,kJ/mol, and 3.05 for wet crustal diabase and $2.4\times10^{-16}$\,Pa$^{-3.5}$s$^{-1}$, 540\,kJ/mol, and 3.5 for dry mantle olivine, respectively; in this context, ``wet'' means that the material is saturated with water, which is already the case for the volumetrically (and thus rheologically) important minerals at very low concentrations under the low pressure conditions of the crust. We allow for the possibility of an incompetent layer separating the crust and mantle segments of the elastic lithosphere but did not actually observe such a configuration in our models. Therefore, the elastic thickness of the lithosphere is tied to a single temperature $T_\mathrm{bd}$ as defined in Eq.~\ref{eq:Tbd} associated with the brittle--ductile transition of the mantle ($\sim1065$\,K in our case), similar to \citet{Rued:etal13b,Rued:etal13a}, whose transition temperature was a few dozen kelvins lower; the depth at which this temperature is reached is taken to represent $z_\mathrm{el}$. In some instances, structures like cold downwellings make it difficult to determine reasonably the local elastic thickness, and it seems safer to discard such regions when computing the global mean; this problem arose in the models with low water content especially in the time interval from $\sim 250$\,My to $\sim 1.5$\,Gy, and therefore the corresponding curves in Figure~\ref{fig:qszl_t}b should be regarded as approximate only in that time interval.\par
The surface heat flow of the reference models (Figure~\ref{fig:qszl_t}) shows a small increase at the earliest stage of evolution before declining smoothly toward present-day values close to 26.5\,mW/m$^2$; the curves are almost the same in all reference models except for that initial phase (Figure~\ref{fig:qszl_t}a). In general, the heat flux does not show large variation in the Reference models: deviations from the global mean by more than 1--2\,mW/m$^2$ are rare. The elastic thicknesses grow almost linearly for most of the time toward a modern value of a bit more than 200\,km, except for some minor variability during the first few hundreds of millions of years; the elastic thicknesses in the Ref-144 pair are some 20\,km larger than those of the Ref-36 pair, but the difference decreases as the planet evolves. All models are broadly consistent with most literature data, especially for the first 1.5\,Gy (Figure~\ref{fig:qszl_t}b); in such comparisons it must be kept in mind that several of the heat flow estimates were made for specific regional structures such as volcanic edifices or graben and that their age is very uncertain.\par
Impact events produce a sudden change in the conditions near the surface in the surroundings of the impact site: they shock-heat the interior beneath the point of impact and trigger melt production, which leads to magmatic activity. In our treatment, we assume that 1.) the erupted material loses heat quickly and can therefore be set to surface temperature; and that 2.) a large volume of ejecta is produced, much of which is deposited as a layer of several kilometers thickness and with surface temperature within the crater. The surface porosity within the crater is assumed to be zero due to melt filling the voids. The heat flux in the affected area, i.e., within the final crater is strongly reduced but returns to higher levels with time, although the levels of undisturbed areas are not reached again; only in the case of the Huygens-sized events is the drop preceded by a spike-like maximum, which in that case is due to the direct heating effect of the impact itself. The temporal heat flux evolutions for any given impact size are quite similar for all three impact times in terms of the amplitude of the disturbance and the duration of its decay and result in almost identical regional present-day heat flows. Moreover, the regional heat flux evolutions of the Isidis-sized and the Utopia-size events at 400\,My are very similar at any given water content, such that the curves for the Utopia-sized models in Figure~\ref{fig:qszl_t}a are fairly representative for both. The reason for the reduction of the heat flow, which is contrary to intuition and to the findings of other authors \citep[e.g.,][]{JHRoAr-Ha12}, is that these large events produce a substantial amount of ejecta and erupted material that is assumed to have surface temperature when being deposited as a layer, whose ejecta component alone can be more than 10\,km thick. This material is represented as a cold influx that shifts isotherms downward and thus causes a very shallow temperature gradient that reduces the heat flow, especially inside the final crater. This effect of ejecta and other shattered material was not taken into consideration in most previous studies. \citet{Rolf:etal17} do not observe the cooling effect in their models of the Moon, because they treat ejecta deposition in a different way and assume that ejecta are absent within the basin; they find a reduced heat flux just beyond the final crater rim, which in their case is due to the reduced thermal conductivity of the relatively thick ejecta deposits there. Moreover, the crust formed in the crater in our models is produced from high-degree melting of a source that was already partly depleted, so that its content in heat-producing elements is several times lower than the average, resulting in a reduced crustal contribution to the surface heat flux. The hot pulse from the mantle diffuses only slowly upward and is then consumed in heating the cool ejecta-thickened crust. The present-day elastic thickness of the lithosphere beneath the crater in models with impacts is usually not significantly different from the global mean of impact-free models, reaching values between 200 and $\sim230$\,km, irrespective of the time of impact; the only stronger deviation from this pattern is the U-144 model pair for an impact at 50\,My model time, in which $z_\mathrm{el}$ reaches values around 250\,km.\par
On a global scale, even large impacts produce only a small instantaneous negative excursion from the values of the Reference models, followed by an exponential-like return toward the heat flux of the corresponding reference model within a few tens of millions of years. The thermal disturbance from a single impact early in the planet's history decays quickly on the timescale of planetary evolution and leaves no distinguishable trace today. In this context it is worth pointing out that a calculation of global mean values by computing the surface heat flow for every single grid column, summing, and dividing by the number of columns as done for the Reference models would be misleading in two-dimensional models with impacts and is therefore omitted here. The reason is that a single local anomaly like an impact site occupies a certain fraction of the domain that is larger than the true fraction of the entire planet that is affected by an impact in three dimensions, because the two-dimensional geometry implicitly extends that anomaly along half a meridian in the third dimension, making its spatial extent potentially much larger than it would be in three dimensions. As a consequence, it contributes more to the mean than it would in three dimensions, thus resulting in an overestimate. Although this applies, in principle, to all major intrinsically three-dimensional structures in two-dimensional models, the effects of structures that are ubiquitous in the model and can make positive or negative contributions to the mean may cancel out reasonably well for the purpose of calculations such as mean global heat flow estimates; this is unfortunately not the case with a single local structure such as an impact. Appendix~\ref{app:2Dgeom} gives a simple estimate of this geometrical error and indicates that the true amplitudes of the impact signal would be one to two orders of magnitude smaller even for large impacts.
\subsubsection{Density and seismic velocities}
\begin{figure}[ht!]
\includegraphics[width=\textwidth]{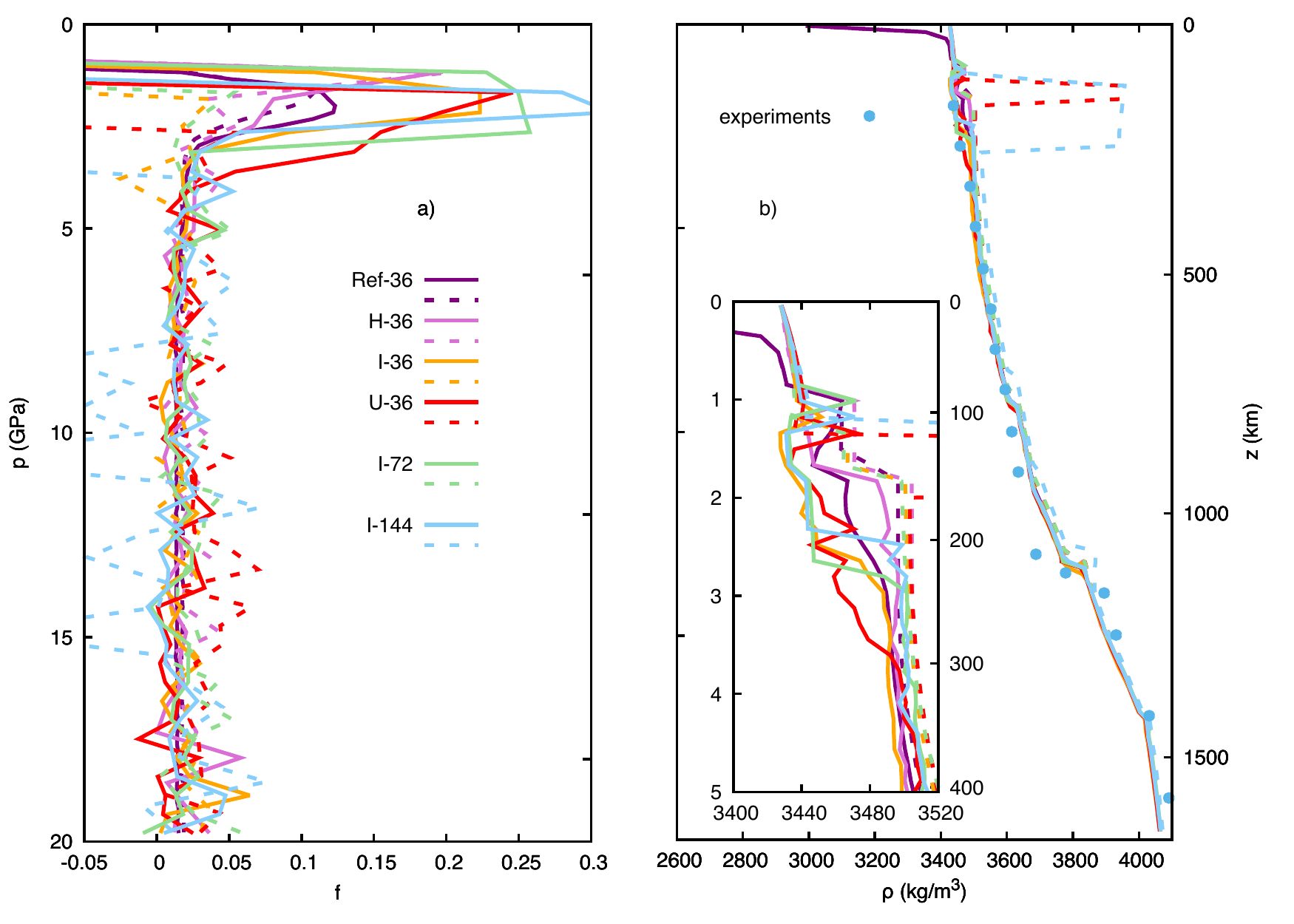}
\caption{Mantle depth (pressure) profiles of depletion (a) and density (b) at 4.4\,Gy for selected models. The profiles of the reference models are laterally averaged profiles, the profiles of the models with impacts were taken at the impact site; the depletion profiles of the latter set were smoothed for clarity. Also shown are experimentally determined densities from \citet{BeFe98} for a composition modeled after \citet{WaDr94}.\label{fig:f-rho_z}}
\end{figure}

Our thermoelastic properties model directly yields the density $\varrho$ of the crust and mantle at different depletions. Subtraction of the laterally averaged density--depth profiles of the T model from the TC model of a model pair indicates that the compositional density difference due to melting is up to 30\,kg/m$^3$; locally, as in impact-generated anomalies with very high depletion, it may exceed 50\,kg/m$^3$. This additional contribution to buoyancy has certain dynamical implications that become especially important on a local scale in the models with impacts, as can be seen by comparing the evolutions of the two models of a given pair. In impacts that are large enough to reach into the mantle, the anomaly created by the strong heating and depletion has a large buoyancy and rises toward the base of the lithosphere, where it tends to spread, as mentioned above (cf. Figure~\ref{fig:isidis}). Contrary to the thermal anomaly, which vanishes as the excess heat diffuses away, the depletion anomaly could only be destroyed by remixing. This does indeed happen in the T models, but the compositional contribution to the buoyancy active in the TC models helps to stabilize the anomaly in those cases. The signature of the impact is clearly visible in present-day depth profiles of depletion of the TC models at the site, where the preserved depletion from the event is still nearly up to three times as high as the average maximum of impact-free models. It is correlated with density anomalies (Figure~\ref{fig:f-rho_z}b), i.e., it has survived 4\,Gy of planetary evolution, whereas it has largely vanished in the T models. The inset figure, which covers the uppermost 400\,km/5\,GPa of the mantle, shows clearly the density deficit of the compositional anomalies of the two larger impact categories relative to the Reference model, whereas the uppermost slice of mantle below the Huygens site happens to be less depleted even than the Reference and is therefore a bit denser. By contrast, the purely thermal models all have a higher and almost identical density in this depth range; the two high-density segments in models U-36/T and U-144/T downward from 100\,km are due to eclogitic crustal material that is beginning to founder. The profiles for the 72 and 144\,ppm water models look very similar and show the same features, except that they seem to reach a bit less deep; the reason is probably that the generally lower viscosity in these models allows density anomalies to rise and spread out laterally more easily. The same pattern is observed in the models with earlier impact times.\par
As the thermoelastic model is based on an equation of state, it also provides the isothermal bulk modulus $K_T$ directly, although it does not yet include the shear modulus and its derivatives. Nonetheless, as a first step toward application to seismic data, we calculate the bulk sound speed in the infinite-frequency approximation,
\begin{equation}
v_\mathrm{B}=\sqrt{\frac{K_S}{\varrho}}=\sqrt{v_\mathrm{P}^2-\frac{4}{3}v_\mathrm{S}^2};
\end{equation}
the adiabatic bulk modulus $K_S$ can be calculated directly from the available thermoelastic properties. This relation permits the direct comparison with data from a seismometer on Mars.\par
\begin{figure}[ht!]
\includegraphics[width=0.5\textwidth]{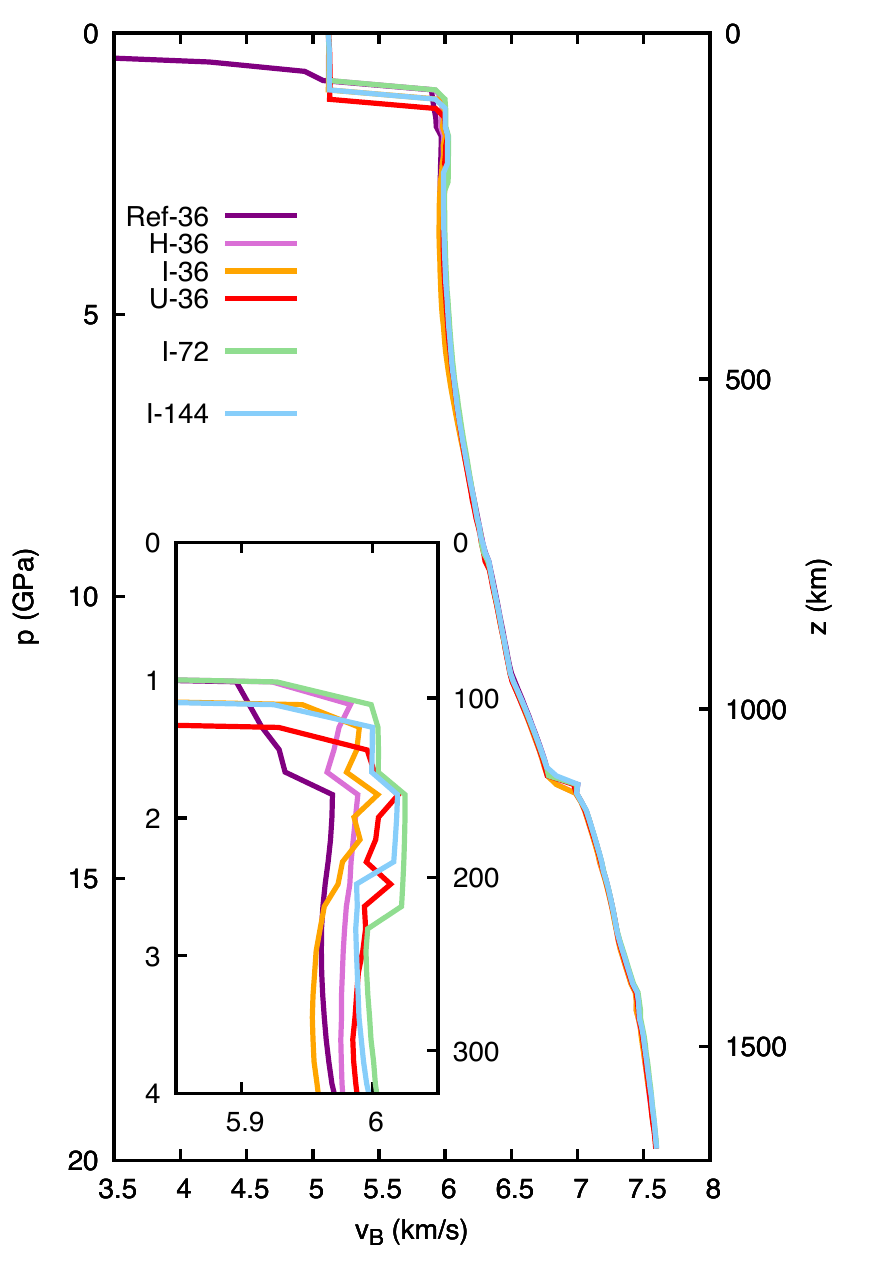}
\caption{Mantle depth (pressure) profile of bulk sound velocity from TC models from the 36\,ppm water series at 4.4\,Gy. The Ref-36 profile is laterally averaged, the profiles of the models with impacts were taken at the impact site.\label{fig:vB_z}}
\end{figure}

Figure~\ref{fig:vB_z} shows representative depth profiles of TC models from the 36\,ppm water series. The profiles for models with impacts correspond to the impact site and show the signature of the remnants of the impact in the topmost mantle.  The magnitude of the seismic anomaly can be assessed by comparison with the laterally averaged profile for model Ref-36, which is virtually identical with the laterally averaged profiles of the impact-affected models; the seismic $v_\mathrm{B}$ anomaly is on the order of a few tens of meters per second. The anomalies in the other model series are of a similar magnitude; as we did not consider possible effects of water due to limited data and the generally low water concentrations in these models, the water-related effects on seismic velocity are not included here, but would likely be negligible in our models. The seismic profile shows a slightly thickened crust at the impact site in models I-36 and U-36, but given the limitations on spatial resolution, the thickening should not be interpreted quantitatively.

\section{Discussion}
The most interesting dynamical feature of the models of this study are certain effects of the compositional contribution to the buoyancy of mantle material on the character of convection and the stability of the lithosphere, the asthenosphere, and local structures therein. \citet{PlBr14} have studied the effect of different compositional density contrasts on convection in a more idealized Mars-like planet with imposed density contrasts and found that the depleted low-density layer is dynamically stable enough to suppress melting and crust formation to a significant extent and thus result in a thinner crust if the density difference between depleted material in the melting zone and fertile rock exceeds a certain value. Our models, in which the density contrast is not imposed but calculated from the compositional changes the mantle material undergoes upon melting, confirm the finding of those authors and suggest that in Mars density contrasts between fertile and depleted mantle rock may indeed reach a magnitude that throttles melt production. Moreover, local features with extreme density contrasts, exemplified here by anomalies created by basin-forming impacts, rise toward the base of the lithosphere, where they spread out under the influence of their own strong buoyancy. The long-lived compositional component of that buoyancy, which is not destroyed by diffusion, makes it possible for them to survive until they are integrated into the lithosphere, facilitating their preservation for the entire remaining history of the planet. \citet{Rolf:etal17} came to a similar conclusion in their models of the Moon.\par
As Mars has been hit by many impacts large enough to penetrate into the asthenosphere or even deeper, impacts should be expected to have produced substantial lateral chemical heterogeneity in the uppermost martian mantle, because they would remelt material in the regular melting zone to different degrees according to their sizes and depths of penetration. The largest events would even affect mantle material that lies below the regular melting region and may therefore be more pristine. The resulting ancient melts from the impact would span a wide range of compositions. Moreover, their source regions in the mantle, which are depleted to different degrees, would themselves form a set of heterogeneous source regions for later volcanism if they become reactivated as melt sources, e.g., by plumes. All this heterogeneity may be reflected in the chemical variability observed in the different types of martian meteorites. Although the impacts we have modeled postdate the supposed time around 4.5\,Ga \citep[e.g.,][]{Borg:etal97,Borg:etal16} at which the major heterogeneity distinguishing the different classes of martian meteorites was established by several hundred millions of years, the mechanism as such could have worked at those earlier times as well; the model pairs with impacts at earlier times indicate that it is viable even at the more vigorous convection when the lithosphere was thinner and the mantle even hotter, and the higher number of giant impacts would offer more opportunities to form the corresponding strong anomalies. Such a scenario would come close to Option~3 of \citet[Fig.~6]{Borg:etal16}, in which a major part of the mantle of the recently accreted and differentiated early Mars is molten during giant impacts and undergoes silicate differentiation, which produces separate shergottite and nakhlite source regions by $\sim 4.5$\,Ga. Apart from these regional scenarios, we note that even regular melting could induce some degree of compositional layering with a density structure that counteracts rehomogenization, especially at higher initial temperatures than assumed here, as shown already before by \citet{PlBr14}. That scenario resembles the one proposed by \citet{JHJones03}, who has suggested that the mantle sources of nakhlites and the more depleted shergottites lie at different depths and are largely isolated from each other. Finally, if a part of the impactor material merges with the mantle target, this effect may permit it to leave its signature in those preserved anomalies, although we did not include the potentially different composition of the impactor in the model.\par
Apart from the obvious crater at the surface, how can old impact sites be detected? The most conspicuous feature of an impact, the crater, is being obliterated with time and may all but disappear. Indeed, even for very large impacts, it is not quite clear how many have occurred in the ancient past of Mars: an analysis by \citet{Frey08} tallied 20 circular depressions with more than 1000\,km diameter that are considered traces of giant impacts, whereas a more recent revision by \citet{FrMa14} came up with 31 candidates. These analyses were based on topography, modeled crustal thicknesses, and surface geology, all of which depend on assumptions or may be affected by later influences such as rebound or the spreading and leveling of crustal thickness undulations. Under such circumstances, additional criteria provided by gravimetry or seismology could, in principle, be helpful.\par
\begin{figure}[ht!]
\includegraphics[width=0.5\textwidth]{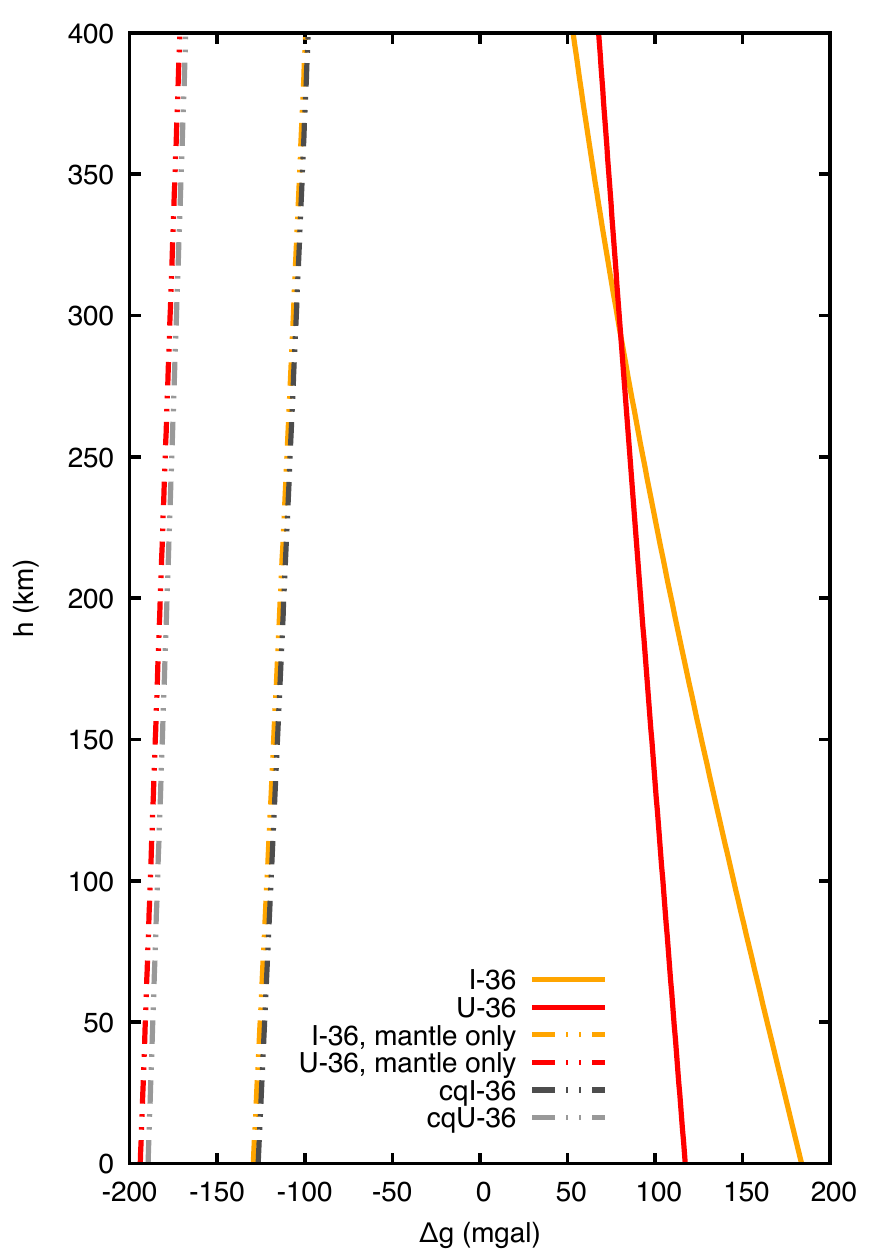}
\caption{Estimated on-axis gravity anomaly at the TC models for I-36 and U-36 as a function of height above the martian surface according to eq.~\ref{eq:dgcyl}. Also shown are the signals of positive crustal thickness anomalies above normal mantle (cqI-36 and cqU-36) that would be almost identical to the impact-generated mantle anomalies in I-36 and U-36, respectively, as discussed in the text.\label{fig:gz_h}}
\end{figure}

Figure~\ref{fig:f-rho_z} shows some representative density profiles that quantify the density contrast that characterizes the self-stabilizing impact-created anomalies, which at that stage is entirely compositional. The present-day density deficit associated with very large impacts such as Isidis or even Utopia reaches its maximum of $\sim 30$\,kg/m$^3$ at a depth of 120 or 220\,km, respectively, and may be overlain by anomalous structures such as a thickened crust or a thin slice of less depleted mantle. Neglecting curvature, we make a rough estimate of the expected gravity anomaly by approximating the deficit mass of the mantle anomaly by a flat, thick cylinder of radius $R_\mathrm{a}$, height $h_\mathrm{a}$, and homogeneous density contrast $\Delta\varrho$ extending downward from a depth $z_\mathrm{a}$. The vertical component of the gravity anomaly on its axis relative to the reference model mantle is given by
\begin{equation}
\Delta g_z=2\pi G_0 \Delta\varrho\left(h_\mathrm{a}+\sqrt{z^2_\mathrm{a}+R^2_\mathrm{a}}-\sqrt{(z_\mathrm{a}+h_\mathrm{a})^2+R^2_\mathrm{a}}\right)\label{eq:dgcyl}
\end{equation}
and can reach values of $-129$ and $-194$\,mgal for I-36/TC ($h_\mathrm{a}=135$\,km, $R_\mathrm{a}\approx \pi R_\mathrm{P}/6$) and U-36/TC ($h_\mathrm{a}=195$\,km, $R_\mathrm{a}\approx \pi R_\mathrm{P}/3$), respectively (Fig.~\ref{fig:gz_h}), assuming $\Delta\varrho=-25$\,kg/m$^3$ and $z_\mathrm{a}=85$\,km in both cases. We find a crust thickened by $\sim 20$\,km in both impact models, but notice that the vertical resolution of our models does not permit a much better determination of the crustal thickness than the grid point spacing ($\sim 13$\,km). To estimate the crustal contribution, we construct a three-cylinder model consisting of a deep cylinder ($\Delta\varrho=-12$ or $-20$\,kg/m$^3$ for Isidis and Utopia) representing the excess crust and two stacked cylinders ($\Delta\varrho=10$ or 250\,kg/m$^3$) each spanning half of the depth range of normal crust and representing the denser, pore-free crust in the final crater; the radius of the crustal anomaly is taken to be that of the final crater. The resulting positive anomalies will result in fairly similar total gravity signals from these two impacts, and they would at any rate be visible from the ground as well as from an orbit at the elevation of the Mars Reconnaissance Orbiter (MRO) ($\sim 400$\,km) (Figure~\ref{fig:gz_h}). However, due to the non-uniqueness of potential fields, such signatures cannot be taken as proof of the presence of an impact-related mantle anomaly. To illustrate this point, one can try to mimic the impact-generated mantle anomalies by crust thickened by a certain amount over an area of the dimensions of our assumed mantle anomaly but overlying an undisturbed mantle. As the gray curves labeled cqI-36 and cqU-36m in Figure~\ref{fig:gz_h} show, thickening by 15 or 22\,km of a crust with $\Delta\varrho=-210$\,kg/m$^3$ would almost perfectly match the I-36/TC and U-36/TC mantle anomalies, respectively. An important implication of this trade-off is that neglecting negative densities anomalies in the mantle such as those caused by ancient basin-forming impacts may result in an overestimate of the crustal thickness. However, we emphasize that this calculation for a thickened post-impact crust is only an example based on the larger impacts in this study; depending on the parameters of the impact, the post-impact crust could also be thinner. Net crustal thinning is often invoked in the interpretation of gravity signatures of impact basins. Therefore, our approach to extract almost all of the melt generated directly by the impact and assume that it contributes to the formation of the post-impact crust only puts an upper bound on the crustal thickness but may be an overestimate. The complex processes occurring during the rebound and collapse of the transient crater, the crystallization of a post-impact magma pond, and the different modes of melt migration under these conditions cannot be modeled within the framework of our global mantle convection models. Dedicated investigations of the impact process itself will be needed to provide a parameterization that allows to determine the fraction of impact-generated melt that actually forms new post-impact crust in a modeling setup like ours. At any rate, present models of the global crustal thickness variation on Mars \citep[e.g.,][]{Zube:etal00,WiZu04} are based on the assumption of a constant mantle density beneath the crust and may therefore produce inaccurate thickness estimates.\par
A seismic survey targeted at a potential impact site will certainly map the crust--mantle boundary and thus resolve deviations in crustal thickness, but the detection of the bulk velocity anomaly in the mantle, which is not expected to exceed 50\,m/s (cf. Figure~\ref{fig:vB_z}), would require a very good coverage of the shallow mantle with seismic rays and a reliable reference model; whether actual P or S wave anomalies are more readily seen is yet to be investigated. Although such observations are in principle possible, the observational framework is not expected to come into existence in the next few decades.\par
The global surface heat flow curves are very similar for all models over almost the entire evolution of the planet, as we stuck to one radionuclide composition; this is in agreement with the findings by \citet{Ples:etal15}, who arrive at similar values for their models with the \citet{WaDr94} concentrations. The heat flow experiment HP$^3$ on the InSight mission, which is set to arrive at Mars in 2018, will hopefully answer the question whether the radionuclide composition by \citet{WaDr94} is correct, but it will not be able to discern between other characteristics of interior models such as viscosity or water content due to the insensitivity of $q_\mathrm{s}$. Estimates of the elastic thickness of the lithosphere and the mean lithospheric heat flow, which are already available, are burdened with more assumptions but provide more means to distinguish between different scenarios, especially with regard to the water content and its effect on convective vigor. We did not consider local heat flow variations in more detail, because they are poorly resolved and depend on a variety of assumptions. For instance, local variations arise in our models as a consequence of crustal temperature modifications after impacts, whereas in the lunar models by \citet{Rolf:etal17}, they are caused by variations in the thickness of ejecta layers with a low thermal conductivity. We agree with those authors in the sense that both of us assigned an insulating effect to ejecta or regolith in general, but contrary to \citet{Rolf:etal17}, we assume that the permanent meteorite flux had produced an insulating global regolith layer from the beginning, which is implemented as a pressure-dependent layer of reduced thermal conductivity and density. Similar to them, we reset the porosity of the surface layer to zero within the final crater by invoking magmatic processes that fill the pore space; this results in positive mass anomalies in the crust within the crater. As it is impossible to quantify plausibly the ratio of intrusive to extrusive magmatism in the aftermath of the impact or the detailed temperature distribution of the ejecta and the newly formed crust in the framework of these global models, we set the temperature of the crater infilling to the surface value as the simplest solution but caution that this results in an underestimate of the post-impact crustal temperature; the crater surface heat flows in Fig.~\ref{fig:qszl_t} should therefore be regarded as lower bounds. At any rate, local variations of conductivity at the surface are not expected to have any substantial effect on the dynamics and processes in the deeper interior, especially in the direct aftermath of the impact, and on the resulting density anomalies.

\section{Conclusions}
The compositional contribution to buoyancy in convection models with melting can produce a stable, depleted layer beneath the lithosphere that acts to reduce melt production and may help to establish separate chemical reservoirs. The same factor appears to be crucial for anchoring highly depleted compositional anomalies, which may for instance be produced by large impacts, beneath the lithosphere and stabilize them for timespans of hundreds of millions of years, allowing them to be permanently included into the growing lithosphere. Such stable anomalies may appear as different reservoirs in the geochemical record.\par
In spite of this mechanism of preservation, anomalies such as those created by giant impacts are heavily modified by mantle convection, and their geophysical signature is expected to degrade with time. As the chemical anomaly entails a density anomaly, it may be possible to detect it in gravity data; by contrast, the prospects for detecting it with seismic means seem rather poor, and the heat flux anomaly directly caused by the impact is nil. Thermal signals from single impacts up to the size considered here do not have a long-lasting, discernible effect on the global heat flow; whether that is different for the cumulative effects of several basin-forming impacts will depend on their magnitudes and timing and will be investigated in a future study.

\appendix
\section{Martian mantle solidus}\label{app:sol}
We have revised the mantle solidus derived in \citet{Rued:etal13a}, which was based on data by \citet{BeHo94a} and \citet{Schm:etal01} from experiments on a K-free martian model peridotite similar to the compositional model from \citet{WaDr94}; we decided not to use the results from \citet{BoDr03} and \citet{AgDr04} although it would be desirable to have more data at higher $p$, because these authors used a material with somewhat different iron and alkali contents. More recently, \citet{Mats:etal13} and \citet{Coll:etal15} performed new melting experiments on martian mantle analogue materials after \citet{WaDr94}, the former also on a K-free composition. As \citet{Coll:etal15} pointed out, the lack of K in the other studies would lead to an increase of the solidus by $\sim$30\,K; this estimate is consistent with the effect of K on the solidus observed in the K-doped peridotite from \citet{WWaTa00}, if the difference in K content is accounted for by a linear dependence. Therefore, we reduced the temperature values measured at upper mantle pressures from the other studies by 30\,K. As \citet{Coll:etal15} did not provide subsolidus data, we extrapolated the melting degree--temperature relations from their experiments linearly at each pressure and weighted the resulting solidus estimate doubly. In view of the high-$p$ values from \citet{Schm:etal01}, it seems that the solidus determined by \citet{Coll:etal15} themselves is unsuited for extrapolation beyond $\sim$3.5\,GPa, and so we decided to fit a selection of all available data for pressures up to the appearance of pv+fp anew, which gave:
\begin{equation}
T_\mathrm{s}=
\begin{cases}
0.118912p^3-6.37695p^2+130.33p+1340.38&\text{for $p<23$\,GPa}\\
62.5p+975&\mathrm{otherwise,}
\end{cases}\label{eq:solidus}
\end{equation}
for $p$ in GPa and $T_\mathrm{s}$ in K. The linear fit to the highest-$p$ points from \citet{Schm:etal01} for the pv+fp stability field was only slightly shifted to avoid a kink at the transition but was otherwise left unchanged from our previous work, because the depressing effect of K on the solidus was observed to disappear at lower-mantle conditions \citep{WWaTa00}; in our work, we rarely approach the solidus at that depth anyway, and pv+fp does not appear at all in the models of this study.\par
Apart from the data for the solidus itself, the work of \citet{Mats:etal13} and \citet{Coll:etal15} was also used to readjust estimates for the temperatures at which certain mineral phases are exhausted during melting, especially for the garnet stability field.\par
\citet{Kief:etal15} discussed the effect of Na, K, and Ca on the martian solidus and proposed shifting the original terrestrial peridotite solidus by \citet{Hirschmann00} down by 35\,K in order to account for that effect. However, \citet[section~4.1]{Katz:etal03} had already pointed out that the original \citet{Hirschmann00} solidus, which was based on intermediately fertile terrestrial compositions, seems to overestimate the solidus temperature as a consequence of the poor coverage of low-degree melting by experimental data and had suggested lowering it by 35\,K, which they found to give a better fit to the experimental data; in other words, their improved fit for terrestrial peridotite is practically identical to the function \citet{Kief:etal15} suggest for martian peridotite. Following the argument of \citet{Kief:etal15} concerning the effect of Na, K, and Ca, which is still valid, then implies that it is the function from \citet{Katz:etal03} that should be reduced by further 35\,K in order to produce a martian solidus.\par
\begin{figure}[ht!]
\includegraphics[width=\textwidth]{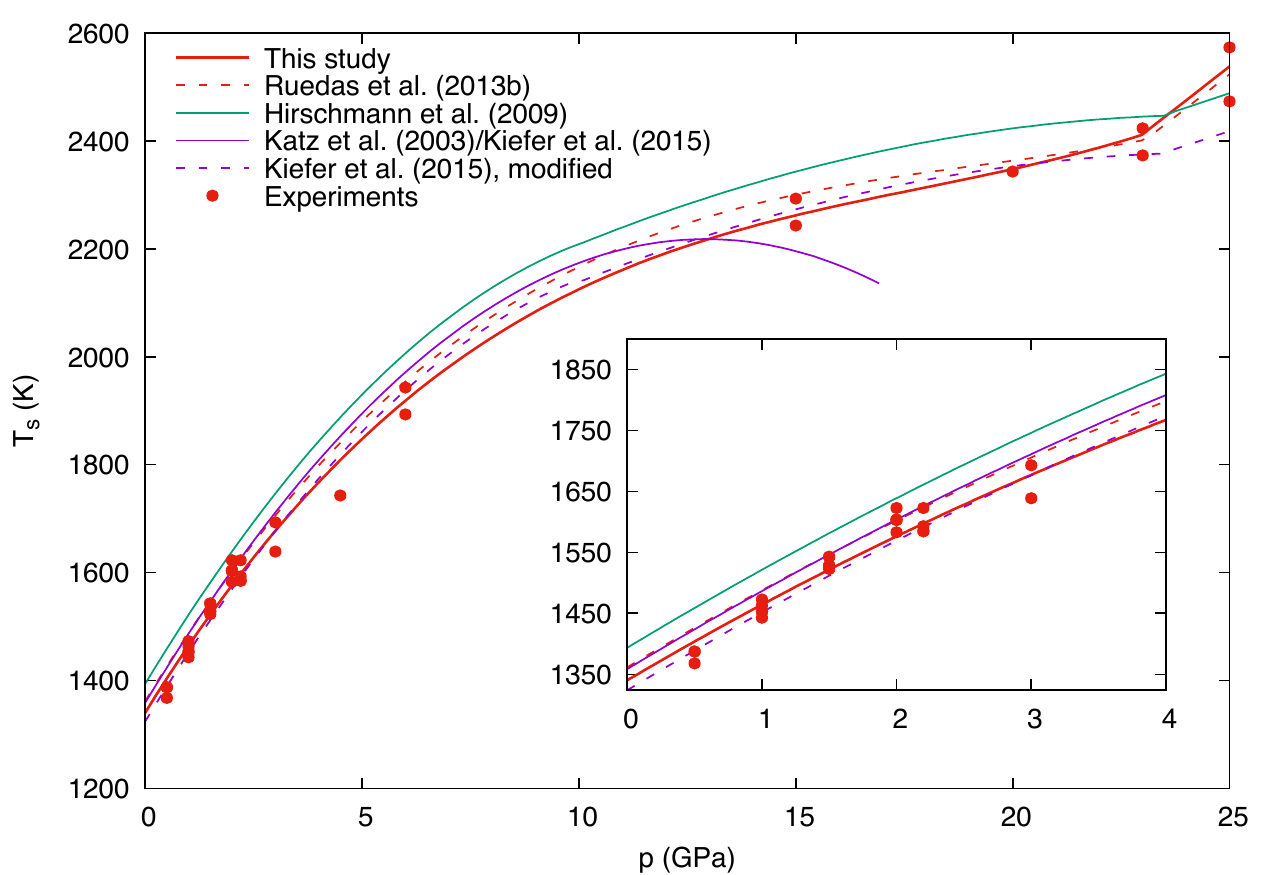}
\caption{Different peridotite dry solidus functions. The experimental data are those used for fitting in this study, i.e., the data from \citet{BeHo94a}, \citet{Schm:etal01}, \citet{Mats:etal13} (all reduced by 30\,K), and \citet{Coll:etal15}.\label{fig:Ts}}
\end{figure}

Figure~\ref{fig:Ts} shows the functions proposed here and demonstrates that a further reduction of the \citet{Kief:etal15} solidus would result in a function that agrees very well with ours especially in the most critical pressure range up to 3\,GPa. We have plotted the solidus from \citet{MMHirs:etal09} instead of the one from \citet{Hirschmann00}, because the former can be used at pressures above 10\,GPa while being identical to the latter below that pressure. The solidus modified from \citet{Kief:etal15} applies the additional shift and uses the \citet{MMHirs:etal09} solidus (shifted down by 70\,K) at $p>10$\,GPa to make it applicable to higher pressures.\par
All of these solidi have used experimental data on nominally anhydrous materials. However, as recently pointed out by \citet{Sara:etal17}, even nominally anhydrous samples that have been dried in order to remove water may still contain traces of it at the level of a few dozen or hundreds of parts per million. Those authors assert, on the basis of experiments of their own that apply a novel technique for monitoring water content, that previous studies systematically underestimated solidus temperatures because of such trace water. The considerations above concerning the shifts of the solidus due to compositional variations remain correct even if their conclusions are confirmed, but the true anhydrous solidus would then have to be determined by an additional correction that would counteract the depression caused by alkalis and iron. Specifically, for a pressure of 1.5\,GPa, at which they conducted their experiments, they inferred a water content of 140\,ppm in the nominally anhydrous samples on which the fit of the \citet{Hirschmann00} solidus was based and deduced that an increase of the solidus temperature by $\sim60$\,K is necessary to correct this error. This would imply that the solidus originally determined by \citet{MMHirs:etal09} for terrestrial anhydrous peridotite would in fact have to be lowered by mere 10\,K to apply to Mars. However, we think that the data to devise a more generally applicable correction are not yet sufficient at this point, and so we stick to the conventional datasets.

\section{Estimate of the geometrical error in 2D global mean calculations}\label{app:2Dgeom}
We consider a single axisymmetric anomaly on the surface of a spherical planet. In two dimensions, such an anomaly would be represented by a cross-section through the center of the sphere and the center of the anomaly, and the anomaly would appear as an arc of length $\theta$, i.e., it occupies a fraction $f_2=\theta/(2\pi)$ of the perimeter of the cross-section. By contrast, in three dimensions the anomaly is a spherical cap that occupies a fraction
\begin{equation}
f_3=\frac{1}{4\pi}\int\limits_0^{2\pi} \int\limits_0^{\frac{\theta}{2}} \sin\theta\,\mathrm{d}\theta \mathrm{d}\varphi=\frac{1}{2}\left(1-\cos\frac{\theta}{2}\right)
\end{equation}
of the total surface of the sphere. Hence, the amplitude marking the impact in the two-dimensional model should be multiplied by a factor
\begin{equation}
\frac{f_3}{f_2}=\frac{\pi}{\theta}\left(1-\cos\frac{\theta}{2}\right),
\end{equation}
with $\theta$ in radians; the discrepancy is worst for small anomalies. For the Huygens, Isidis, and Utopia events, we have angles $\theta$ of 1.26\textdegree, 3.64\textdegree, and 9.1\textdegree, respectively, and so the correction factors would be 0.0086, 0.025, and 0.062, respectively.

\nocite{AbSt64,AsMe76,BlWo03b,BJWoBl14,Debye12,DuAn89,Fornberg98,Gruneisen12,
Gruneisen26,ItSt92,JHJones95,Mame:etal09,StBu90,StDa04,Taur:etal98,Toplis05}
\nocite{Koga:etal03,Auba:etal04,Haur:etal06,Tenn:etal09,OLear:etal10,Nove:etal14b,Tenn:etal12a,
vKanP:etal10,BrKo90,Lang:etal92,Blun:etal95,Harlow97,Haur:etal94b,Cham:etal02,Broo:etal03,
Dyge:etal14,Land:etal01,vWest:etal01a,BJWoBl02,RaGr74,Shannon93,BJWood:etal99,Mibe:etal06b,
JChen:etal02,Ohta:etal00,CoWo04,BoCa:etal00,YHZhao:etal09,Stacey98,Leon:etal11,Mack:etal98,
ZMJin:etal01,Hofmeister99,HiKo03,HeWi92,Berryman95}
\newpage
\subsection*{Notation}
The subscript ``imp'' indicates a property of the impactor throughout the text.\\[1ex]
\begin{longtable}{rl}
\endfirsthead
\caption{(cont'd.)}\\
\endhead
\endlastfoot
$a$, $b$&scaling and shift parameters for $p_\mathrm{s}$ after \citet{Ruedas17}\\
$B$&dislocation creep pre-factor\\
$c_p$&heat capacity\\
$C_i$&concentration of $i$\\
$D_\mathrm{f}$&final crater diameter\\
$D_\mathrm{imp}$&impactor diameter\\
$D_\mathrm{tr}$&transient crater diameter\\
$D_\mathrm{s2c}\approx 5.6$\,km&simple/complex transition crater diameter (Mars)\\
$f$&melting degree/depletion\\
$G_0=6.67408\times 10^{-11}$\,m$^3$/(kg\,s$^2$)&Newton's constant of gravitation\\
$g=3.72$\,m/s$^2$&surface gravity acceleration (Mars)\\
$\Delta g_z$&vertical component of gravity anomaly\\
$K_S$&adiabatic bulk modulus\\
$n$&decay exponent for $p_\mathrm{s}$ after \citet{Ruedas17}\\
$p_\mathrm{IM}$&shock pressure from impedance-match solution\\
$p_\mathrm{l}$&lithostatic pressure\\
$p_\mathrm{s}$&peak shock pressure\\
$Q$&activation enthalpy\\
$q_\mathrm{s}$&surface heat flow\\
$R_\mathrm{gas}=8.3144598$\,J/(mol\,K)&universal gas constant\\
$R_\mathrm{P}=3389.5$\,km&planetary radius (Mars)\\
$r_\mathrm{infl}$&radius of isobaric core defined by inflection point of $p_\mathrm{s}$\\
$r_\mathrm{PVM}$&radius of isobaric core after \citet{Pier:etal97}\\
$Ra_0$, $Ra_\mathrm{eff}$&initial, effective Rayleigh number\\
$S_\mathrm{h}$&Hugoniot slope\\
$T$&temperature\\
$T_\mathrm{pot}=1700$\,K&initial potential temperature\\
$T_\mathrm{s}$&solidus temperature\\
$u_\mathrm{IM}$&particle velocity from impedance-match solution\\
$v_\mathrm{B}$&speed of sound\\
$v_\mathrm{imp}=9.6$\,km/s, $v_{z,\mathrm{imp}}=6.8$\,km/s&impact velocity and its vertical component (Mars)\\
$v_\mathrm{P}$, $v_\mathrm{S}$&P-, S-wave velocity\\
$z_\mathrm{el}$&elastic thickness of the lithosphere\\
$z_\mathrm{ic}$&depth to center of isobaric core\\
$\dot\varepsilon=10^{-17}$\,s$^{-1}$&strain rate\\
$\eta_\mathrm{mean}$&volumetric mean of viscosity, nondimensionalized\\
$\varphi$&porosity\\
$\varphi_\mathrm{r}=0.007$&threshold porosity for melt extraction\\
$\gamma$&Gr\"uneisen parameter\\
$\nu$&exponent of creep law\\
$\varrho$&density\\
$\sigma_\mathrm{y}=15$\,MPa&yield stress\\
 \end{longtable}

\subsection*{Acknowledgments}
Anne Hofmeister and Zhicheng Jing kindly gave additional explanations of various aspects of their respective work. We thank Martin Knapmeyer for helpful remarks on some seismological aspects and Sabrina Schwinger for helping to clarify a question concerning the water content of the crust. The comments by Gregor Golabek and an anonymous referee are gratefully acknowledged. TR was supported by grant Ru 1839/1-1 from the Deutsche Forschungsgemeinschaft (DFG), with additional initial funding from the Helmholtz Alliance project ``Planetary evolution and life''. DB was supported by the DFG (SFB-TRR 170). This is TRR~170 publication no.~28. The numerical calculations were carried out on the computational resource ForHLR~II at the Steinbuch Centre for Computing, Karlsruhe Institute of Technology, funded by the Ministry of Science, Research and the Arts Baden-Württemberg and DFG. Some earlier code development first applied in this study benefitted from support by the NASA Planetary Geology and Geophysics Program through grant NNX11AC62G and the use of computing resources provided by the NASA High-End Computing (HEC) Program through the NASA Center for Climate Simulation (NCCS) at Goddard Space Flight Center, award SMD-11-2549. TR is grateful to Sean Solomon for having made this support from NASA possible. The data used in this investigation are presented in this paper and the Supporting Information; data sets used in plots and figures can be obtained from Figshare at \url{https://doi.org/10.6084/m9.figshare.c.3777686}.


\begin{thebibliography}{139}
\providecommand{\natexlab}[1]{#1}
\expandafter\ifx\csname urlstyle\endcsname\relax
  \providecommand{\doi}[1]{doi:\discretionary{}{}{}#1}\else
  \providecommand{\doi}{doi:\discretionary{}{}{}\begingroup
  \urlstyle{rm}\Url}\fi

\bibitem[{\textit{Abramov et~al.}(2012)\textit{Abramov, Wong, and
  Kring}}]{Abra:etal12}
Abramov, O., S.~M. Wong, and D.~A. Kring (2012), Differential melt scaling for
  oblique impacts on terrestrial planets, \textit{Icarus}, \textit{218}(2),
  906--916, \doi{10.1016/j.icarus.2011.12.022}.

\bibitem[{\textit{Abramowitz and Stegun}(1964)}]{AbSt64}
Abramowitz, M., and I.~A. Stegun (Eds.) (1964), \textit{Handbook of
  Mathematical Functions}, U.S. National Bureau of Standards/Dover Publications, New York.

\bibitem[{\textit{Agee and Draper}(2004)}]{AgDr04}
Agee, C.~B., and D.~S. Draper (2004), {Experimental constraints on the origin
  of Martian meteorites and the composition of the Martian mantle},
  \textit{Earth Planet. Sci. Lett.}, \textit{224}(3--4), 415--429.

\bibitem[{\textit{Ashcroft and Mermin}(1976)}]{AsMe76}
Ashcroft, N.~W., and N.~D. Mermin (1976), \textit{Solid State Physics},
  Saunders College Publishing.

\bibitem[{\textit{Aubaud et~al.}(2004)\textit{Aubaud, Hauri, and
  Hirschmann}}]{Auba:etal04}
Aubaud, C., E.~H. Hauri, and M.~M. Hirschmann (2004), Hydrogen partition
  coefficients between nominally anhydrous minerals and basaltic melts,
  \textit{Geophys. Res. Lett.}, \textit{31}, L20611,
  \doi{10.1029/2004GL021341}.

\bibitem[{\textit{Balog et~al.}(2001)\textit{Balog, Secco, and
  Rubie}}]{Balo:etal01}
Balog, P.~S., R.~A. Secco, and D.~C. Rubie (2001), {Density measurements of
  liquids at high pressure: Modifications to the sink/float method by using
  composite spheres, and application to Fe-10\,wt\%S}, \textit{High Press. Res.},
  \textit{21}(5), 237--261, \doi{10.1080/08957950108201026}.

\bibitem[{\textit{Balog et~al.}(2003)\textit{Balog, Secco, Rubie, and
  Frost}}]{Balo:etal03}
Balog, P.~S., R.~A. Secco, D.~C. Rubie, and D.~J. Frost (2003), {Equation of
  state of liquid Fe-10\,wt\% S: Implications for the metallic cores of
  planetary bodies}, \textit{J. Geophys. Res.}, \textit{108}(B2), 2124,
  \doi{10.1029/2001JB001646}.

\bibitem[{\textit{Barnett and Nimmo}(2002)}]{BaNi02}
Barnett, D.~N., and F.~Nimmo (2002), {Strength of faults on Mars from MOLA
  topography}, \textit{Icarus}, \textit{157}(1), 34--42,
  \doi{10.1006/icar.2002.6817}.

\bibitem[{\textit{Belleguic et~al.}(2005)\textit{Belleguic, Lognonn{\'e}, and
  Wieczorek}}]{Bell:etal05}
Belleguic, V., P.~Lognonn{\'e}, and M.~Wieczorek (2005), {Constraints on the
  Martian lithosphere from gravity and topography data}, \textit{J. Geophys.
  Res.}, \textit{110}, E11005, \doi{10.1029/2005JE002437}.

\bibitem[{\textit{Berryman}(1995)}]{Berryman95}
Berryman, J.~G., Mixture theories for rock properties, in \textit{Rock Physics
  \& Phase Relations --- A Handbook of Physical Constants}, \textit{AGU
  Reference Shelf}, vol.~3, edited by T.~J. Ahrens, pp. 205--228, American
  Geophysical Union, Washington, D.C., 1995.
  
\bibitem[{\textit{Bertka and Fei}(1998)}]{BeFe98}
Bertka, C.~M., and Y.~Fei (1998), {Density profile of an SNC model Martian
  interior and the moment-of-inertia factor of Mars}, \textit{Earth Planet.
  Sci. Lett.}, \textit{157}(1--2), 79--88, \doi{10.1016/S0012-821X(98)00030-2}.

\bibitem[{\textit{Bertka and Holloway}(1994)}]{BeHo94a}
Bertka, C.~M., and J.~R. Holloway (1994), {Anhydrous partial melting of an
  iron-rich mantle I: subsolidus phase assemblages and partial melting phase
  relations at 10 to 30\,kbar}, \textit{Contrib. Mineral. Petrol.},
  \textit{115}(3), 313--322.

\bibitem[{\textit{Bjorkman and Holsapple}(1987)}]{BjHo87}
Bjorkman, M.~D., and K.~A. Holsapple (1987), Velocity scaling impact melt
  volume, \textit{Int. J. Impact Engng.}, \textit{5}(1--4), 155--163,
  \doi{10.1016/0734-743X(87)90035-2}.

\bibitem[{\textit{Blundy and Wood}(2003)}]{BlWo03b}
Blundy, J., and B.~Wood (2003), Mineral--melt partitioning of uranium, thorium
  and their daughters, in \textit{Uranium-series Geochemistry}, edited by
  B.~Bourdon, G.~M. Henderson, C.~C. Lundstrom, and S.~P. Turner, no.~52 in
  Rev. Mineral. Geochem., pp. 59--123, Mineralogical Society of America,
  Washington, D.C.

\bibitem[{\textit{Blundy et~al.}(1995)\textit{Blundy, Falloon, Wood, and
  Dalton}}]{Blun:etal95}
Blundy, J.~D., T.~J. Falloon, B.~J. Wood, and J.~A. Dalton (1995), Sodium
  partitioning between clinopyroxene and silicate melts, \textit{J. Geophys.
  Res.}, \textit{100}(B8), 15,501--15,515.

\bibitem[{\textit{Bolfan-Casanova et~al.}(2000)\textit{Bolfan-Casanova,
  Keppler, and Rubie}}]{BoCa:etal00}
Bolfan-Casanova, N., H.~Keppler, and D.~C. Rubie (2000), {Water partitioning
  between nominally anhydrous minerals in the MgO--SiO\textsubscript{2}--H\textsubscript{2}O
  system up to 24\,GPa: implications for the distribution of water in the
  Earth's mantle}, \textit{Earth Planet. Sci. Lett.}, \textit{182}, 209--221.

\bibitem[{\textit{Borg et~al.}(1997)\textit{Borg, Nyquist, Taylor, Weismann,
  and Shih}}]{Borg:etal97}
Borg, L., L.~E. Nyquist, L.~A. Taylor, H.~Weismann, and C.-Y. Shih (1997),
  {Constraints on Martian differentiation processes from Rb--Sr and Sm--Nd
  isotopic analyses of the basaltic shergottite QUE 94201}, \textit{Geochim.
  Cosmochim. Acta}, \textit{61}(22), 4915--4931.

\bibitem[{\textit{Borg and Draper}(2003)}]{BoDr03}
Borg, L.~E., and D.~S. Draper (2003), {A petrogenetic model for the origin and
  compositional variation of the martian basaltic meteorites},
  \textit{Meteorit. Planet. Sci.}, \textit{38}(12), 1713--1731.

\bibitem[{\textit{Borg et~al.}(2016)\textit{Borg, Brennecka, and
  Symes}}]{Borg:etal16}
Borg, L.~E., G.~A. Brennecka, and S.~J.~K. Symes (2016), {Accretion timescale
  and impact history of Mars deduced from the isotopic systematics of martian
  meteorites}, \textit{Geochim. Cosmochim. Acta}, \textit{175}, 150--167,
  \doi{10.1016/j.gca.2015.12.002}.

\bibitem[{\textit{Breuer et~al.}(2016)\textit{Breuer, Plesa, Tosi, and
  Grott}}]{Breu:etal16}
Breuer, D., A.-C. Plesa, N.~Tosi, and M.~Grott (2016), {Water in the Martian
  interior---The geodynamical perspective}, \textit{Meteorit. Planet. Sci.},
  \textit{51}(11), 1959--1992, \doi{10.1111/maps.12727}.

\bibitem[{\textit{Brey and K{\"o}hler}(1990)}]{BrKo90}
Brey, G.~P., and T.~K{\"o}hler (1990), {Geothermobarometry in four-phase
  lherzolites II. New thermobarometers, and practical assessment of existing
  thermobarometers}, \textit{J. Petrol.}, \textit{31}(6), 1353--1378,
  \doi{10.1093/petrology/31.6.1353}.

\bibitem[{\textit{Brooker et~al.}(2003)\textit{Brooker, Du, Blundy, Kelley,
  Allan, Wood, Chamorro, Wartho, and Purton}}]{Broo:etal03}
Brooker, R.~A., Z.~Du, J.~D. Blundy, S.~P. Kelley, N.~L. Allan, B.~J. Wood,
  E.~M. Chamorro, J.-A. Wartho, and J.~A. Purton (2003), The `zero charge'
  partitioning behaviour of noble gases during mantle melting, \textit{Nature},
  \textit{423}, 738--741, \doi{10.1038/nature01708}.

\bibitem[{\textit{Chamorro et~al.}(2002)\textit{Chamorro, Brooker, Wartho,
  Wood, Kelley, and Blundy}}]{Cham:etal02}
Chamorro, E.~M., R.~A. Brooker, J.-A. Wartho, B.~J. Wood, S.~P. Kelley, and
  J.~D. Blundy (2002), Ar and K partitioning between clinopyroxene and silicate
  melt to 8\,GPa, \textit{Geochim. Cosmochim. Acta}, \textit{66}(3), 507--519.

\bibitem[{\textit{Chen et~al.}(2002)\textit{Chen, Inoue, Yurimoto, and
  Weidner}}]{JChen:etal02}
Chen, J., T.~Inoue, H.~Yurimoto, and D.~Weidner (2002), {Effect of water on
  olivine--wadsleyite phase boundary in the (Mg,Fe)\textsubscript{2}SiO\textsubscript{4}
  system}, \textit{Geophys. Res. Lett.}, \textit{29}(18), 22.

\bibitem[{\textit{Collinet et~al.}(2015)\textit{Collinet, M{\'e}dard, Charlier,
  Vander~Auwera, and Grove}}]{Coll:etal15}
Collinet, M., E.~M{\'e}dard, B.~Charlier, J.~Vander~Auwera, and T.~L. Grove
  (2015), {Melting of the primitive martian mantle at 0.5--2.2 GPa and the
  origin of basalts and alkaline rocks on Mars}, \textit{Earth Planet. Sci.
  Lett.}, \textit{427}, 83--94, \doi{10.1016/j.epsl.2015.06.056}.

\bibitem[{\textit{Corgne and Wood}(2004)}]{CoWo04}
Corgne, A., and B.~J. Wood (2004), {Trace element partitioning between
  majoritic garnet and silicate melt at 25\,GPa}, \textit{Phys. Earth Planet.
  Inter.}, \textit{143--144}, 407--419, \doi{10.1016/j.pepi.2003.08.012}.

\bibitem[{\textit{Croft}(1980)}]{Croft80}
Croft, S.~K. (1980), {Cratering flow fields: Implications for the excavation
  and transient expansion stages of crater formation}, \textit{Lunar Planet.
  Sci. Proc.}, \textit{XI}, 2347--2378.

\bibitem[{\textit{Debye}(1912)}]{Debye12}
Debye, P. (1912), {Zur Theorie der spezifischen W{\"a}rme}, \textit{Ann.
  Phys.}, \textit{IV/39}, 789--839.

\bibitem[{\textit{Duffy and Anderson}(1989)}]{DuAn89}
Duffy, T.~S., and D.~L. Anderson (1989), Seismic velocities in mantle minerals
  and the mineralogy of the upper mantle, \textit{J. Geophys. Res.},
  \textit{94}(B2), 1895--1912.

\bibitem[{\textit{Dygert et~al.}(2014)\textit{Dygert, Liang, Sun, and
  Hess}}]{Dyge:etal14}
Dygert, N., Y.~Liang, C.~Sun, and P.~Hess (2014), {An experimental study of
  trace element partitioning between augite and Fe-rich basalts},
  \textit{Geochim. Cosmochim. Acta}, \textit{132}, 170--186,
  \doi{10.1016/j.gca.2014.01.042, 10.1016/j.gca.2014.09.013}, corrigendum in
  149, 281--283.

\bibitem[{\textit{Fern{\'a}ndez et~al.}(2015)\textit{Fern{\'a}ndez, Li, Howell,
  and Woodney}}]{Fern:etal15}
Fern{\'a}ndez, Y.~R., J.-Y. Li, E.~S. Howell, and L.~M. Woodney (2015),
  Asteroids and comets, in \textit{Physics of Terrestrial Planets and Moons},
  \textit{Treatise on Geophysics}, vol.~10, edited by T.~Spohn, 2nd ed., chap.
  10.15, pp. 487--528, Elsevier, \doi{10.1016/B978-0-444-53802-4.00184-6}.

\bibitem[{\textit{Fornberg}(1998)}]{Fornberg98}
Fornberg, B. (1998), Calculation of weights in finite difference formulas,
  \textit{SIAM Rev.}, \textit{40}(3), 685--691.

\bibitem[{\textit{Frey}(2008)}]{Frey08}
Frey, H. (2008), {Ages of very large impact basins on Mars: Implications for
  the late heavy bombardment in the inner solar system}, \textit{Geophys. Res.
  Lett.}, \textit{35}, L13203, \doi{10.1029/2008GL033515}.

\bibitem[{\textit{Frey and Mannoia}(2014)}]{FrMa14}
Frey, H.~V., and L.~M. Mannoia (2014), {A revised, rated and dated inventory of
  very large candidate impact basins on Mars}, \textit{Lunar Planet. Sci.},
  \textit{45}, 1892.

\bibitem[{\textit{Gault and Heitowit}(1963)}]{GaHe63}
Gault, D.~E., and E.~D. Heitowit (1963), The partition of energy for
  hypervelocity impact craters formed in rock, in \textit{Proc. 6th
  Hypervelocity Impact Symp.}, vol.~2, pp. 419--456, U.S. Army, U.S. Air Force,
  U.S. Navy, Cleveland, Ohio.

\bibitem[{\textit{Gault and Wedekind}(1978)}]{GaWe78}
Gault, D.~E., and J.~A. Wedekind (1978), Experimental studies of oblique
  impact, \textit{Proc. Lunar Planet. Sci. Conf.}, \textit{9}, 3843--3875.

\bibitem[{\textit{Golabek et~al.}(2011)\textit{Golabek, Keller, Gerya, Zhu,
  Tackley, and Connolly}}]{Gola:etal11}
Golabek, G.~J., T.~Keller, T.~V. Gerya, G.~Zhu, P.~J. Tackley, and J.~A.~D.
  Connolly (2011), {Origin of the martian dichotomy and Tharsis from a giant
  impact causing massive magmatism}, \textit{Icarus}, \textit{215}(1),
  346--357, \doi{10.1016/j.icarus.2011.06.012}.

\bibitem[{\textit{Grott and Breuer}(2008)}]{GrBr08a}
Grott, M., and D.~Breuer (2008), The evolution of the martian elastic
  lithosphere and implications for crustal and mantle rheology,
  \textit{Icarus}, \textit{193}(2), 503--515.

\bibitem[{\textit{Gr{\"u}neisen}(1912)}]{Gruneisen12}
Gr{\"u}neisen, E. (1912), {Theorie des festen Zustandes einatomiger Elemente},
  \textit{Ann. Phys.}, \textit{39}, 257--306.

\bibitem[{\textit{Gr{\"u}neisen}(1926)}]{Gruneisen26}
Gr{\"u}neisen, E. (1926), Zustand des festen K{\"o}rpers, in \textit{Thermische
  Eigenschaften der Stoffe}, \textit{Handbuch der Physik}, vol.~10, edited by
  C.~Drucker, E.~Gr{\"u}neisen, P.~Kohnstamm, F.~K{\"o}rber, K.~Scheel,
  E.~Schr{\"o}dinger, F.~Simon, J.~D. van~der Waals, Jr., and F.~Henning,
  chap.~1, pp. 1--59, Springer, Heidelberg.

\bibitem[{\textit{Harlow}(1997)}]{Harlow97}
Harlow, G.~E. (1997), {K in clinopyroxene at high pressure and temperature: An
  experimental study}, \textit{Amer. Mineral.}, \textit{82}(3--4), 259--269.

\bibitem[{\textit{Hauck and Phillips}(2002)}]{HaPh02}
Hauck, S.~A., and R.~J. Phillips (2002), {Thermal and crustal evolution of
  Mars}, \textit{J. Geophys. Res.}, \textit{107}(E7), 5052,
  \doi{10.1029/2001JE001801}.

\bibitem[{\textit{Hauri et~al.}(1994)\textit{Hauri, Wagner, and
  Grove}}]{Haur:etal94b}
Hauri, E.~H., T.~P. Wagner, and T.~L. Grove (1994), {Experimental and natural
  partitioning of Th, U, Pb and other trace elements between garnet,
  clinopyroxene and basaltic melts}, \textit{Chem. Geol.}, \textit{117},
  149--166.

\bibitem[{\textit{Hauri et~al.}(2006)\textit{Hauri, Gaetani, and
  Green}}]{Haur:etal06}
Hauri, E.~H., G.~A. Gaetani, and T.~H. Green (2006), {Partitioning of water
  during melting of the Earth's upper mantle at H\textsubscript{2}O-undersaturated
  conditions}, \textit{Earth Planet. Sci. Lett.}, \textit{248}(3--4), 715--734.

\bibitem[{\textit{Head and Wilson}(1992)}]{HeWi92}
Head, J.~W., and L.~Wilson, {Magma reservoirs and neutral buoyancy zones on
  Venus: Implications for the formation and evolution of volcanic landforms},
  \textit{J. Geophys. Res.}, \textit{97}(E3), 3877--3903,
  \doi{10.1029/92JE00053}, 1992.
  
\bibitem[{\textit{Hernlund and Tackley}(2008)}]{HeTa08}
Hernlund, J.~W., and P.~J. Tackley (2008), Modeling mantle convection in the
  spherical annulus, \textit{Phys. Earth Planet. Inter.}, \textit{171}(1--4),
  48--54, \doi{10.1016/j.pepi.2008.07.037}.

\bibitem[{\textit{Hirschmann}(2000)}]{Hirschmann00}
Hirschmann, M.~M. (2000), {Mantle solidus: Experimental constraints and the
  effects of peridotite composition}, \textit{Geochem. Geophys. Geosyst.},
  \textit{1}, 1042, \doi{10.1029/2000GC000070}.

\bibitem[{\textit{Hirschmann et~al.}(2009)\textit{Hirschmann, Tenner, Aubaud,
  and Withers}}]{MMHirs:etal09}
Hirschmann, M.~M., T.~Tenner, C.~Aubaud, and A.~C. Withers (2009), {Dehydration
  melting of nominally anhydrous mantle: The primacy of partitioning},
  \textit{Phys. Earth Planet. Inter.}, \textit{176}(1--2), 54--68.

\bibitem[{\textit{Hirth and Kohlstedt}(2003)}]{HiKo03}
Hirth, G., and D.~L. Kohlstedt, {Rheology of the upper mantle and the mantle
  wedge: A view from the experimentalists}, in \textit{Inside the Subduction
  Factory}, \textit{AGU Geophysical Monograph}, vol. 138, edited by J.~Eiler,
  pp. 83--105, American Geophysical Union, Washington, D.C., 2003.
  
\bibitem[{\textit{Hofmeister}(1999)}]{Hofmeister99}
Hofmeister, A.~M., Mantle values of thermal conductivity and the geotherm from
  phonon lifetimes, \textit{Science}, \textit{283}, 1699--1706, 1999.
  
\bibitem[{\textit{Hofmeister and Branlund}(2015)}]{HoBr15}
Hofmeister, A.~M., and J.~M. Branlund (2015), Thermal conductivity of the
  earth, in \textit{Mineral Physics}, \textit{Treatise on Geophysics}, vol.~2,
  edited by G.~D. Price and L.~Stixrude, 2nd ed., chap. 2.23, pp. 583--608,
  Elsevier, \doi{10.1016/B978-0-444-53802-4.00047-6}.

\bibitem[{\textit{Ita and Stixrude}(1992)}]{ItSt92}
Ita, J., and L.~Stixrude (1992), Petrology, elasticity, and composition of the
  mantle transition zone, \textit{J. Geophys. Res.}, \textit{97}(B5),
  6849--6866.

\bibitem[{\textit{Ivanov}(2001)}]{Ivanov01}
Ivanov, B.~A. (2001), {Mars/Moon cratering rate ratio estimates}, \textit{Space
  Sci. Rev.}, \textit{96}(1--4), 87--104, \doi{10.1023/A:1011941121102}.

\bibitem[{\textit{Jin et~al.}(2001)\textit{Jin, Zhang, Green, and
  Jin}}]{ZMJin:etal01}
Jin, Z.-M., J.~Zhang, H.~W. Green, II, and S.~Jin, {Eclogite rheology:
  Implications for subducted lithosphere}, \textit{Geology}, \textit{29}(8),
  667--670, 2001.
  
\bibitem[{\textit{Jing and Karato}(2009)}]{ZJiKa09}
Jing, Z., and S.-i. Karato (2009), The density of volatile bearing melts in the
  earth's deep mantle: The role of chemical composition, \textit{Chem. Geol.},
  \textit{262}(1--2), 100--107.

\bibitem[{\textit{Jing and Karato}(2011)}]{JiKa11}
Jing, Z., and S.-i. Karato (2011), {A new approach to the equation of state of
  silicate melts: An application of the theory of hard sphere mixtures},
  \textit{Geochim. Cosmochim. Acta}, \textit{75}(22), 6780--6802,
  \doi{10.1016/j.gca.2011.09.004}.

\bibitem[{\textit{Jing and Karato}(2012)}]{JiKa12}
Jing, Z., and S.-i. Karato (2012), {Effect of H\textsubscript{2}O on the density of
  silicate melts at high pressures: Static experiments and the application of a
  modified hard-sphere model of equation of state}, \textit{Geochim. Cosmochim.
  Acta}, \textit{85}, 357--372, \doi{10.1016/j.gca.2012.03.001}.

\bibitem[{\textit{Johnson et~al.}(2014)\textit{Johnson, Bowling, and
  Melosh}}]{BCJohn:etal14}
Johnson, B.~C., T.~J. Bowling, and H.~J. Melosh (2014), Jetting during vertical
  impacts of spherical projectiles, \textit{Icarus}, \textit{238}, 13--22,
  \doi{10.1016/j.icarus.2014.05.003}.

\bibitem[{\textit{Jones}(1995)}]{JHJones95}
Jones, J.~H. (1995), Experimental trace element partitioning, in \textit{Rock
  Physics and Phase Relations --- A Handbook of Physical Constants},
  \textit{AGU Reference Shelf}, vol.~3, edited by T.~J. Ahrens, pp. 73--104,
  American Geophysical Union, Washington, D.C.

\bibitem[{\textit{Jones}(2003)}]{JHJones03}
Jones, J.~H. (2003), {Constraints on the structure of the martian interior
  determined from the chemical and isotopic systematics of SNC meteorites},
  \textit{Meteorit. Planet. Sci.}, \textit{38}(12), 1807--1814.

\bibitem[{\textit{Kaiura and Toguri}(1979)}]{KaTo79}
Kaiura, G.~H., and J.~M. Toguri (1979), {Densities of the molten FeS,
  FeS--Cu\textsubscript{2}S and Fe--S--O systems~-- Utilizing a bottom-balance
  Archimedean technique}, \textit{Can. Metall. Q.}, \textit{18}(2), 155--164,
  \doi{10.1179/cmq.1979.18.2.155}.

\bibitem[{\textit{Katz et~al.}(2003)\textit{Katz, Spiegelman, and
  Langmuir}}]{Katz:etal03}
Katz, R.~F., M.~Spiegelman, and C.~Langmuir (2003), A new parameterization of
  hydrous mantle melting, \textit{Geochem. Geophys. Geosyst.}, \textit{4}(9),
  1073, \doi{10.1029/2002GC000433}.

\bibitem[{\textit{Keller and Tackley}(2009)}]{KeTa09}
Keller, T., and P.~J. Tackley (2009), {Towards self-consistent modeling of the
  martian dichotomy: The influence of one-ridge convection on crustal thickness
  distribution}, \textit{Icarus}, \textit{202}(2), 429--443,
  \doi{10.1016/j.icarus.2009.03.029}.

\bibitem[{\textit{Khan and Connolly}(2008)}]{KhCo08}
Khan, A., and J.~A.~D. Connolly (2008), {Constraining the composition and
  thermal state of Mars from inversion of geophysical data}, \textit{J.
  Geophys. Res.}, \textit{113}(E7), E07003, \doi{10.1029/2007JE002996}.

\bibitem[{\textit{Kiefer}(2003)}]{Kiefer03}
Kiefer, W.~S. (2003), {Melting in the martian mantle: Shergottite formation and
  implications for present-day mantle convection on Mars}, \textit{Meteorit.
  Planet. Sci.}, \textit{38}(12), 1815--1832.

\bibitem[{\textit{Kiefer et~al.}(2015)\textit{Kiefer, Filiberto, Sandu, and
  Li}}]{Kief:etal15}
Kiefer, W.~S., J.~Filiberto, C.~Sandu, and Q.~Li (2015), {The effects of mantle
  composition on the peridotite solidus: Implications for the magmatic history
  of Mars}, \textit{Geochim. Cosmochim. Acta}, \textit{162}, 247--258,
  \doi{10.1016/j.gca.2015.02.010}.

\bibitem[{\textit{Koga et~al.}(2003)\textit{Koga, Hauri, Hirschmann, and
  Bell}}]{Koga:etal03}
Koga, K., E.~Hauri, M.~M. Hirschmann, and D.~Bell (2003), {Hydrogen
  concentration analyses using SIMS and FTIR: Comparison and calibration for
  nominally anhydrous minerals}, \textit{Geochem. Geophys. Geosyst.},
  \textit{4}(2), \doi{10.1029/2002GC000378}.

\bibitem[{\textit{Konopliv et~al.}(2011)\textit{Konopliv, Asmar, Folkner,
  Karatekin, Nunes, Smrekar, Yoder, and Zuber}}]{Kono:etal11}
Konopliv, A.~S., S.~W. Asmar, W.~M. Folkner, {\"O}.~Karatekin, D.~C. Nunes,
  S.~E. Smrekar, C.~F. Yoder, and M.~T. Zuber (2011), {Mars high resolution
  gravity fields from MRO, Mars seasonal gravity, and other dynamical
  parameters}, \textit{Icarus}, \textit{211}(1), 401--428,
  \doi{10.1016/j.icarus.2010.10.004}.

\bibitem[{\textit{Landwehr et~al.}(2001)\textit{Landwehr, Blundy,
  Chamorro-Perez, Hill, and Wood}}]{Land:etal01}
Landwehr, D., J.~Blundy, E.~M. Chamorro-Perez, E.~Hill, and B.~Wood (2001),
  U-series disequilibria generated by partial melting of spinel lherzolite,
  \textit{Earth Planet. Sci. Lett.}, \textit{188}(3--4), 329--348,
  \doi{10.1016/S0012-821X(01)00328-4}.

\bibitem[{\textit{Langmuir et~al.}(1992)\textit{Langmuir, Klein, and
  Plank}}]{Lang:etal92}
Langmuir, C.~H., E.~M. Klein, and T.~Plank (1992), Petrological systematics of
  mid-ocean ridge basalts: Constraints on melt generation beneath ocean ridges,
  in \textit{Mantle Flow and Melt Generation at Mid-Ocean Ridges}, edited by
  J.~{Phipps Morgan}, D.~K. Blackman, and J.~M. Sinton, no.~71 in Geophysical
  Monographs, pp. 183--280, American Geophysical Union, Washington, D.C.,
  \doi{10.1029/GM071p0183}.

\bibitem[{\textit{Leone et~al.}(2011)\textit{Leone, Wilson, and
  Davies}}]{Leon:etal11}
Leone, G., L.~Wilson, and A.~G. Davies, {The geothermal gradient of Io:
  Consequences for lithosphere structure and volcanic eruptive activity},
  \textit{Icarus}, \textit{211}(1), 623--635,
  \doi{10.1016/j.icarus.2010.10.016}, 2011.
  
\bibitem[{\textit{Li and Kiefer}(2007)}]{QLiKi07a}
Li, Q., and W.~S. Kiefer (2007), {Mantle convection and magma production on
  present-day Mars: Effects of temperature-dependent rheology},
  \textit{Geophys. Res. Lett.}, \textit{34}, L16203,
  \doi{10.1029/2007GL030544}.

\bibitem[{\textit{Mackwell et~al.}(1998)\textit{Mackwell, Zimmerman, and
  Kohlstedt}}]{Mack:etal98}
Mackwell, S.~J., M.~E. Zimmerman, and D.~L. Kohlstedt, {High-temperature
  deformation of dry diabase with application to tectonics on Venus},
  \textit{J. Geophys. Res.}, \textit{103}(B1), 975--984, 1998.
  
\bibitem[{\textit{Mamedov et~al.}(2009)\textit{Mamedov, Eser, Ko{\c c}, and
  Askerov}}]{Mame:etal09}
Mamedov, B.~A., E.~Eser, H.~Ko{\c c}, and I.~M. Askerov (2009), {Accurate
  evaluation of the specific heat capacity of solids and its application to MgO
  and ZnO crystals}, \textit{Int. J. Thermophys.}, \textit{30}(3), 1048--1054.

\bibitem[{\textit{Matsukage et~al.}(2013)\textit{Matsukage, Nagayo, Whitaker,
  Takahashi, and Kawasaki}}]{Mats:etal13}
Matsukage, K.~N., Y.~Nagayo, M.~L. Whitaker, E.~Takahashi, and T.~Kawasaki
  (2013), {Melting of the Martian mantle from 1.0 to 4.5\,GPa}, \textit{J.
  Miner. Petr. Sci.}, \textit{108}(4), 201--214, \doi{10.2465/jmps.120820}.

\bibitem[{\textit{Maxwell}(1977)}]{Maxwell77}
Maxwell, D.~E. (1977), {Simple Z model of cratering, ejection, and the
  overturned flap}, in \textit{Impact and explosion cratering}, edited by D.~J.
  Roddy, R.~O. Pepin, and R.~B. Merrill, pp. 1003--1008, Pergamon Press, New
  York.

\bibitem[{\textit{McCubbin et~al.}(2010)\textit{McCubbin, Smirnov, Nekvasil,
  Wang, Hauri, and Lindsley}}]{McCu:etal10}
McCubbin, F.~M., A.~Smirnov, H.~Nekvasil, J.~Wang, E.~Hauri, and D.~H. Lindsley
  (2010), {Hydrous magmatism on Mars: A source for water for the surface and
  subsurface during the Amazonian}, \textit{Earth Planet. Sci. Lett.},
  \textit{292}(1--2), 132--138.

\bibitem[{\textit{McCubbin et~al.}(2012)\textit{McCubbin, Hauri, Elardo,
  Vander~Kaaden, Wang, and Shearer}}]{McCu:etal12a}
McCubbin, F.~M., E.~H. Hauri, S.~M. Elardo, K.~E. Vander~Kaaden, J.~Wang, and
  C.~K. Shearer, Jr. (2012), Hydrous melting of the martian mantle produced
  both depleted and enriched shergottites, \textit{Geology}, \textit{40}(8),
  683--686, \doi{10.1130/G33242.1}.

\bibitem[{\textit{McCubbin et~al.}(2016)\textit{McCubbin, Boyce, Srinivasan,
  Santos, Elardo, Filiberto, Steele, and Shearer}}]{McCu:etal16a}
McCubbin, F.~M., J.~W. Boyce, P.~Srinivasan, A.~R. Santos, S.~M. Elardo,
  J.~Filiberto, A.~Steele, and C.~K. Shearer (2016), {Heterogeneous
  distribution of H\textsubscript{2}O in the Martian interior: Implications for the
  abundance of H\textsubscript{2}O in depleted and enriched mantle sources},
  \textit{Meteorit. Planet. Sci.}, \textit{51}(11), 2036--2060,
  \doi{10.1111/maps.12639}.

\bibitem[{\textit{McGovern et~al.}(2002)\textit{McGovern, Solomon, Smith,
  Zuber, Simons, Wieczorek, Phillips, Neumann, Aharonson, and
  Head}}]{McGo:etal02}
McGovern, P.~J., S.~C. Solomon, D.~E. Smith, M.~T. Zuber, M.~Simons, M.~A.
  Wieczorek, R.~J. Phillips, G.~A. Neumann, O.~Aharonson, and J.~W. Head
  (2002), {Localized gravity/topography admittance and correlation spectra on
  Mars: Implications for regional and global evolution}, \textit{J. Geophys.
  Res.}, \textit{107}(E12), 5136, \doi{10.1029/2002JE001854}, correction in
  vol.~109, E07007, 10.1029/2004JE002286 (2004).

\bibitem[{\textit{Melosh}(1989)}]{Melosh89}
Melosh, H.~J. (1989), \textit{Impact cratering: a geologic process}, no.~11 in
  Oxford Monographs on Geology and Geophysics, ix+245 pp., Oxford University
  Press.

\bibitem[{\textit{Melosh}(2011)}]{Melosh11}
Melosh, H.~J. (2011), \textit{Planetary Surface Processes}, no.~13 in Cambridge
  Planetary Science, Cambridge University Press.

\bibitem[{\textit{Mibe et~al.}(2006)\textit{Mibe, Orihashi, Nakai, and
  Fujii}}]{Mibe:etal06b}
Mibe, K., Y.~Orihashi, S.~Nakai, and T.~Fujii (2006), Element partitioning
  between transition-zone minerals and ultramafic melt under hydrous
  conditions, \textit{Geophys. Res. Lett.}, \textit{33}, L16307,
  \doi{10.1029/2006GL026999}.

\bibitem[{\textit{Monteux and Arkani-Hamed}(2016)}]{MoAr-Ha16}
Monteux, J., and J.~Arkani-Hamed (2016), Scaling laws of impact induced shock
  pressure and particle velocity in planetary mantle, \textit{Icarus},
  \textit{264}, 246--256, \doi{10.1016/j.icarus.2015.09.040}.

\bibitem[{\textit{Morschhauser et~al.}(2011)\textit{Morschhauser, Grott, and
  Breuer}}]{Mors:etal11}
Morschhauser, A., M.~Grott, and D.~Breuer (2011), {Crustal recycling, mantle
  dehydration, and the thermal evolution of Mars}, \textit{Icarus},
  \textit{212}(2), 541--558, \doi{10.1016/j.icarus.2010.12.028}.

\bibitem[{\textit{Nimmo et~al.}(2004)\textit{Nimmo, Price, Brodholt, and
  Gubbins}}]{Nimm:etal04}
Nimmo, F., G.~D. Price, J.~Brodholt, and D.~Gubbins (2004), The influence of
  potassium on core and geodynamo evolution, \textit{Geophys. J. Int.},
  \textit{156}(2), 363--376.

\bibitem[{\textit{Nishida et~al.}(2011)\textit{Nishida, Ohtani, Urakawa,
  Suzuki, Sakamaki, Terasaki, and Katayama}}]{Nish:etal11}
Nishida, K., E.~Ohtani, S.~Urakawa, A.~Suzuki, T.~Sakamaki, H.~Terasaki, and
  Y.~Katayama (2011), {Density measurement of liquid FeS at high pressures
  using synchrotron X-ray absorption}, \textit{Amer. Mineral.},
  \textit{96}(5--6), 864--868.

\bibitem[{\textit{Novella et~al.}(2014)\textit{Novella, Frost, Hauri, Bureau,
  Raepsaet, and Roberge}}]{Nove:etal14b}
Novella, D., D.~J. Frost, E.~H. Hauri, H.~Bureau, C.~Raepsaet, and M.~Roberge
  (2014), {The distribution of H\textsubscript{2}O between silicate melt and nominally
  anhydrous peridotite and the onset of hydrous melting in the deep upper
  mantle}, \textit{Earth Planet. Sci. Lett.}, \textit{400}, 1--13,
  \doi{10.1016/j.epsl.2014.05.006}.

\bibitem[{\textit{Ogawa and Yanagisawa}(2011)}]{OgYa11}
Ogawa, M., and T.~Yanagisawa (2011), Numerical models of martian mantle
  evolution induced by magmatism and solid-state convection beneath stagnant
  lithosphere, \textit{J. Geophys. Res.}, \textit{116}, E08008,
  \doi{10.1029/2010JE003777}.

\bibitem[{\textit{Ogawa and Yanagisawa}(2012)}]{OgYa12}
Ogawa, M., and T.~Yanagisawa (2012), Two-dimensional numerical studies on the
  effects of water on martian mantle evolution induced by magmatism and
  solid-state mantle convection, \textit{J. Geophys. Res.}, \textit{117},
  E06004, \doi{10.1029/2012JE004054}.

\bibitem[{\textit{Ohtani et~al.}(2000)\textit{Ohtani, Mizobata, and
  Yurimoto}}]{Ohta:etal00}
Ohtani, E., H.~Mizobata, and H.~Yurimoto (2000), {Stability of dense hydrous
  magnesium silicate phases in the systems
  Mg\textsubscript{2}SiO\textsubscript{4}--H\textsubscript{2}O and MgSiO\textsubscript{3}--H\textsubscript{2}O at
  pressures up to 27\,GPa}, \textit{Phys. Chem. Min.}, \textit{27}(8),
  533--544, \doi{10.1007/s002690000097}.

\bibitem[{\textit{O'Keefe and Ahrens}(1985)}]{OKeAh85}
O'Keefe, J.~D., and T.~J. Ahrens (1985), Impact and explosion crater ejecta,
  fragment size, and velocity, \textit{Icarus}, \textit{62}(2), 328--338,
  \doi{10.1016/0019-1035(85)90128-9}.

\bibitem[{\textit{O'Keefe and Ahrens}(1987)}]{OKeAh87}
O'Keefe, J.~D., and T.~J. Ahrens (1987), The size distributions of fragments
  ejected at a given velocity from impact craters, \textit{Int. J. Impact
  Engng.}, \textit{5}(1--4), 493--499, \doi{10.1016/0734-743X(87)90064-9}.

\bibitem[{\textit{O'Leary et~al.}(2010)\textit{O'Leary, Gaetani, and
  Hauri}}]{OLear:etal10}
O'Leary, J.~A., G.~A. Gaetani, and E.~H. Hauri (2010), {The effect of
  tetrahedral Al\textsuperscript{3+} on the partitioning of water between clinopyroxene
  and silicate melt}, \textit{Earth Planet. Sci. Lett.}, \textit{297}(1--2),
  111--120.

\bibitem[{\textit{Phillips et~al.}(2008)\textit{Phillips, Zuber, Smrekar,
  Mellon, Head, Tanaka, Putzig, Milkovich, Campbell, Plaut, Safaeinili, Seu,
  Biccari, Carter, Picardi, Orosei, Mohit, Heggy, Zurek, Egan, Giacomoni,
  Russo, Cutigni, Pettinelli, Holt, Leuschen, and Marinangeli}}]{RJPhil:etal08}
Phillips, R.~J., M.~T. Zuber, S.~E. Smrekar, M.~T. Mellon, J.~W. Head, K.~L.
  Tanaka, N.~E. Putzig, S.~M. Milkovich, B.~A. Campbell, J.~J. Plaut,
  A.~Safaeinili, R.~Seu, D.~Biccari, L.~M. Carter, G.~Picardi, R.~Orosei, P.~S.
  Mohit, E.~Heggy, R.~W. Zurek, A.~F. Egan, E.~Giacomoni, F.~Russo, M.~Cutigni,
  E.~Pettinelli, J.~W. Holt, C.~J. Leuschen, and L.~Marinangeli (2008), {Mars
  north polar deposits: Stratigraphy, age, and geodynamical response},
  \textit{Science}, \textit{320}(5880), 1182--1185,
  \doi{10.1126/science.1157546}.

\bibitem[{\textit{Pierazzo et~al.}(1997)\textit{Pierazzo, Vickery, and
  Melosh}}]{Pier:etal97}
Pierazzo, E., A.~M. Vickery, and H.~J. Melosh (1997), A reevaluation of impact
  melt production, \textit{Icarus}, \textit{127}(2), 408--423,
  \doi{10.1006/icar.1997.5713}.

\bibitem[{\textit{Plesa and Breuer}(2014)}]{PlBr14}
Plesa, A.-C., and D.~Breuer (2014), {Partial melting in one-plate planets:
  Implications for thermo-chemical and atmospheric evolution}, \textit{Planet.
  Space Sci.}, \textit{98}, 50--65, \doi{10.1016/j.pss.2013.10.007}.

\bibitem[{\textit{Plesa et~al.}(2015)\textit{Plesa, Tosi, Grott, and
  Breuer}}]{Ples:etal15}
Plesa, A.-C., N.~Tosi, M.~Grott, and D.~Breuer (2015), {Thermal evolution and
  Urey ratio of Mars}, \textit{J. Geophys. Res.}, \textit{120}(5), 995--1010,
  \doi{10.1002/2014JE004748}.

\bibitem[{\textit{Poirier}(2000)}]{Poirier00}
Poirier, J.-P. (2000), \textit{Introduction to the Physics of the Earth's
  Interior}, 2 ed., Cambridge University Press.

\bibitem[{\textit{Potter et~al.}(2015)\textit{Potter, Kring, and
  Collins}}]{Pott:etal15}
Potter, R. W.~K., D.~A. Kring, and G.~S. Collins (2015), Scaling of basin-sized
  impacts and the influence of target temperature, in \textit{Large Meteorite
  Impacts and Planetary Evolution V}, edited by G.~R. Osinski and D.~A. Kring,
  no. 518 in Special Papers, pp. 99--113, Geological Society of America,
  \doi{10.1130/2015.2518(06)}.

\bibitem[{\textit{R{\aa}heim and Green}(1974)}]{RaGr74}
R{\aa}heim, A., and D.~H. Green (1974), {Experimental determination of the
  temperature and pressure dependence of the Fe-Mg partition coefficient for
  coexisting garnet and clinopyroxene}, \textit{Contrib. Mineral. Petrol.},
  \textit{48}(3), 179--203.

\bibitem[{\textit{Reese and Solomatov}(2006)}]{ReSo06}
Reese, C.~C., and V.~S. Solomatov (2006), Fluid dynamics of local martian magma
  oceans, \textit{Icarus}, \textit{184}(1), 102--120.

\bibitem[{\textit{Reese et~al.}(2002)\textit{Reese, Solomatov, and
  Baumgardner}}]{Rees:etal02}
Reese, C.~C., V.~S. Solomatov, and J.~R. Baumgardner (2002), Survival of
  impact-induced thermal anomalies in the martian mantle, \textit{J. Geophys.
  Res.}, \textit{107}(E10), 5082, \doi{10.1029/2000JE001474}.

\bibitem[{\textit{Reese et~al.}(2004)\textit{Reese, Solomatov, Baumgardner,
  Stegman, and Vezolainen}}]{Rees:etal04}
Reese, C.~C., V.~S. Solomatov, J.~R. Baumgardner, D.~R. Stegman, and A.~V.
  Vezolainen (2004), {Magmatic evolution of impact-induced Martian mantle
  plumes and the origin of Tharsis}, \textit{J. Geophys. Res.}, \textit{109},
  E08009, \doi{10.1029/2003JE002222}.

\bibitem[{\textit{Richardson et~al.}(2007)\textit{Richardson, Melosh, Lisse,
  and Carcich}}]{JERich:etal07}
Richardson, J.~E., H.~J. Melosh, C.~M. Lisse, and B.~Carcich (2007), {A
  ballistics analysis of the Deep Impact ejecta plume: Determining Comet Tempel
  1's gravity, mass, and density}, \textit{Icarus}, \textit{190}(2), 357--390,
  \doi{10.1016/j.icarus.2007.08.001}.

\bibitem[{\textit{Ritzer and Hauck}(2009)}]{RiHa09}
Ritzer, A., and S.~A. Hauck, II (2009), {Lithospheric structure and tectonics
  at Isidis Planitia, Mars}, \textit{Icarus}, \textit{201}(2), 528--539,
  \doi{10.1016/j.icarus.2009.01.025}.

\bibitem[{\textit{Rivoldini et~al.}(2011)\textit{Rivoldini, Van~Hoolst,
  Verhoeven, Mocquet, and Dehant}}]{Rivo:etal11}
Rivoldini, A., T.~Van~Hoolst, O.~Verhoeven, A.~Mocquet, and V.~Dehant (2011),
  {Geodesy constraints on the interior structure and composition of Mars},
  \textit{Icarus}, \textit{213}(2), 451--472,
  \doi{10.1016/j.icarus.2011.03.024}.

\bibitem[{\textit{Robbins and Hynek}(2012)}]{RoHy12b}
Robbins, S.~J., and B.~M. Hynek (2012), {A new global database of Mars impact
  craters $\geq$1 km: 2. Global crater properties and regional variations of
  the simple-to-complex transition diameter}, \textit{J. Geophys. Res.},
  \textit{117}(E6), E06001, \doi{10.1029/2011JE003967}.

\bibitem[{\textit{Roberts and Arkani-Hamed}(2012)}]{JHRoAr-Ha12}
Roberts, J.~H., and J.~Arkani-Hamed (2012), {Impact-induced mantle dynamics on
  Mars}, \textit{Icarus}, \textit{218}(1), 278--289,
  \doi{10.1016/j.icarus.2011.11.038}.

\bibitem[{\textit{Roberts et~al.}(2009)\textit{Roberts, Lillis, and
  Manga}}]{JHRobe:etal09}
Roberts, J.~H., R.~J. Lillis, and M.~Manga (2009), {Giant impacts on early Mars
  and the cessation of the Martian dynamo}, \textit{J. Geophys. Res.},
  \textit{114}, E04009, \doi{10.1029/2008JE003287}.

\bibitem[{\textit{Rolf et~al.}(2017)\textit{Rolf, Zhu, W{\"u}nnemann, and
  Werner}}]{Rolf:etal17}
Rolf, T., M.-H. Zhu, K.~W{\"u}nnemann, and S.~C. Werner (2017), The role of
  impact bombardment history in lunar evolution, \textit{Icarus}, \textit{286},
  138--152, \doi{10.1016/j.icarus.2016.10.007}.

\bibitem[{\textit{Ruedas}(2017)}]{Ruedas17}
Ruedas, T. (2017), Globally smooth approximations for shock pressure decay in
  impacts, \textit{Icarus}, \textit{289}, 22--33,
  \doi{10.1016/j.icarus.2017.02.008}.

\bibitem[{\textit{Ruedas et~al.}(2013{\natexlab{a}})\textit{Ruedas, Tackley,
  and Solomon}}]{Rued:etal13b}
Ruedas, T., P.~J. Tackley, and S.~C. Solomon (2013{\natexlab{a}}), {Thermal and
  compositional evolution of the martian mantle: Effect of water},
  \textit{Phys. Earth Planet. Inter.}, \textit{220}, 50--72,
  \doi{10.1016/j.pepi.2013.04.006}.

\bibitem[{\textit{Ruedas et~al.}(2013{\natexlab{b}})\textit{Ruedas, Tackley,
  and Solomon}}]{Rued:etal13a}
Ruedas, T., P.~J. Tackley, and S.~C. Solomon (2013{\natexlab{b}}), {Thermal and
  compositional evolution of the martian mantle: Effects of phase transitions
  and melting}, \textit{Phys. Earth Planet. Inter.}, \textit{216}, 32--58,
  \doi{10.1016/j.pepi.2012.12.002}.

\bibitem[{\textit{Sanloup et~al.}(2000)\textit{Sanloup, Guyot, Gillet, Fiquet,
  Mezouar, and Martinez}}]{Sanl:etal00}
Sanloup, C., F.~Guyot, P.~Gillet, G.~Fiquet, M.~Mezouar, and I.~Martinez
  (2000), {Density measurements of liquid Fe-S alloys at high-pressure},
  \textit{Geophys. Res. Lett.}, \textit{27}(6), 811--814,
  \doi{10.1029/1999GL008431}.

\bibitem[{\textit{Sarafian et~al.}(2017)\textit{Sarafian, Gaetani, Hauri, and
  Sarafian}}]{Sara:etal17}
Sarafian, E., G.~A. Gaetani, E.~H. Hauri, and A.~R. Sarafian (2017),
  Experimental constraints on the damp peridotite solidus and oceanic mantle
  potential temperature, \textit{Science}, \textit{355}(6328), 942--945,
  \doi{10.1126/science.aaj2165}.

\bibitem[{\textit{Schmerr et~al.}(2001)\textit{Schmerr, Fei, and
  Bertka}}]{Schm:etal01}
Schmerr, N.~C., Y.~Fei, and C.~Bertka (2001), {Extending the solidus for a
  model iron-rich martian mantle composition to 25\,GPa}, \textit{Lunar Planet.
  Sci.}, \textit{XXXII}, Abstract 1157.

\bibitem[{\textit{Schott et~al.}(2001)\textit{Schott, van~den Berg, and
  Yuen}}]{BScho:etal01}
Schott, B., A.~P. van~den Berg, and D.~A. Yuen (2001), Focussed time-dependent
  martian volcanism from chemical differentiation coupled with variable thermal
  conductivity, \textit{Geophys. Res. Lett.}, \textit{28}(22), 4271--4274.

\bibitem[{\textit{Schultz and Glicken}(1979)}]{PHScGl79}
Schultz, P.~H., and H.~Glicken (1979), {Impact crater and basin control of
  igneous processes on Mars}, \textit{J. Geophys. Res.}, \textit{84}(B14),
  8033--8047, \doi{10.1029/JB084iB14p08033}.

\bibitem[{\textit{Shannon}(1993)}]{Shannon93}
Shannon, R.~D. (1993), Dielectric polarizabilities of ions in oxides and
  fluorides, \textit{J. Appl. Phys.}, \textit{73}(1), 348--366,
  \doi{10.1063/1.353856}.

\bibitem[{\textit{Shaw}(2000)}]{DMShaw00}
Shaw, D.~M. (2000), Continuous (dynamic) melting theory revisited, \textit{Can.
  Mineral.}, \textit{38}(5), 1041--1063.

\bibitem[{\textit{Spudis}(1993)}]{Spudis93}
Spudis, P.~D. (1993), \textit{The Geology of Multi-Ring Impact Basins},
  Cambridge University Press.

\bibitem[{\textit{Stacey}(1998)}]{Stacey98}
Stacey, F.~D., Thermoelasticity of a mineral composite and a reconsideration of
  lower mantle properties, \textit{Phys. Earth Planet. Inter.}, \textit{106},
  219--236, 1998.
  
\bibitem[{\textit{Stacey and Davis}(2004)}]{StDa04}
Stacey, F.~D., and P.~M. Davis (2004), High pressure equations of state with
  applications to the lower mantle and core, \textit{Phys. Earth Planet.
  Inter.}, \textit{142}(3--4), 137--184, \doi{10.1016/j.pepi.2004.02.003}.

\bibitem[{\textit{Stixrude and Bukowinski}(1990)}]{StBu90}
Stixrude, L., and M.~S.~T. Bukowinski (1990), Fundamental thermodynamic
  relations and silicate melting with implications for the constitution of
  \ensuremath{\mathrm{D}^{\prime\prime}}, \textit{J. Geophys. Res.}, \textit{95}(B12), 19,311--19,325.

\bibitem[{\textit{Tackley}(1996)}]{Tackley96a}
Tackley, P.~J. (1996), Effects of strongly variable viscosity on
  three-dimensional compressible convection in planetary mantles, \textit{J.
  Geophys. Res.}, \textit{101}(B2), 3311--3332.

\bibitem[{\textit{Tackley}(2008)}]{Tackley08}
Tackley, P.~J. (2008), Modelling compressible mantle convection with large
  viscosity contrasts in a three-dimensional spherical shell using the yin-yang
  grid, \textit{Phys. Earth Planet. Inter.}, \textit{171}(1--4), 7--18.

\bibitem[{\textit{Tauber et~al.}(1978)\textit{Tauber, Kirk, and
  Gault}}]{Taub:etal78}
Tauber, M.~E., D.~B. Kirk, and D.~E. Gault (1978), {An analytic study of impact
  ejecta trajectories in the atmospheres of Venus, Mars, and Earth},
  \textit{Icarus}, \textit{33}, 529--536, \doi{10.1016/0019-1035(78)90188-4}.

\bibitem[{\textit{Taura et~al.}(1998)\textit{Taura, Yurimoto, Kurita, and
  Sueno}}]{Taur:etal98}
Taura, H., H.~Yurimoto, K.~Kurita, and S.~Sueno (1998), Pressure dependence on
  partition coefficients for trace elements between olivine and the coexisting
  melts, \textit{Phys. Chem. Min.}, \textit{25}(7), 469--484,
  \doi{10.1007/s002690050138}.

\bibitem[{\textit{Taylor and McLennan}(2009)}]{SRTaMcLe09}
Taylor, S.~R., and S.~M. McLennan (2009), \textit{Planetary Crusts}, 400 pp.,
  Cambridge University Press.

\bibitem[{\textit{Tenner et~al.}(2009)\textit{Tenner, Hirschmann, Withers, and
  Hervig}}]{Tenn:etal09}
Tenner, T.~J., M.~M. Hirschmann, A.~C. Withers, and R.~L. Hervig (2009),
  {Hydrogen partitioning between nominally anhydrous upper mantle minerals and
  melt between 3 and 5\,GPa and applications to hydrous peridotite partial
  melting}, \textit{Chem. Geol.}, \textit{262}(1--2), 42--56.

\bibitem[{\textit{Tenner et~al.}(2012)\textit{Tenner, Hirschmann, Withers, and
  Ardia}}]{Tenn:etal12a}
Tenner, T.~J., M.~M. Hirschmann, A.~C. Withers, and P.~Ardia (2012),
  {H\textsubscript{2}O storage capacity of olivine and low-Ca pyroxene from 10 to
  13\,GPa: consequences for dehydration melting above the transition zone},
  \textit{Contrib. Mineral. Petrol.}, \textit{163}(2), 297--316,
  \doi{10.1007/s00410-011-0675-7}, erratum in \textit{Contrib. Mineral.
  Petrol.} 163(2), p. 317--318, doi:10.1007/s00410-011-0684-6.

\bibitem[{\textit{Toplis}(2005)}]{Toplis05}
Toplis, M.~J. (2005), The thermodynamics of iron and magnesium partitioning
  between olivine and liquid: criteria for assessing and predicting equilibrium
  in natural and experimental systems, \textit{Contrib. Mineral. Petrol.},
  \textit{149}(1), 22--39.

\bibitem[{\textit{van Kan~Parker et~al.}(2010)\textit{van Kan~Parker,
  Liebscher, Frei, van Sijl, van Westrenen, Blundy, and Franz}}]{vKanP:etal10}
van Kan~Parker, M., A.~Liebscher, D.~Frei, J.~van Sijl, W.~van Westrenen,
  J.~Blundy, and G.~Franz (2010), {Experimental and computational study of
  trace element distribution between orthopyroxene and anhydrous silicate melt:
  substitution mechanisms and the effect of iron}, \textit{Contrib. Mineral.
  Petrol.}, \textit{159}(4), 459--473, \doi{10.1007/s00410-009-0435-0}.

\bibitem[{\textit{van Westrenen et~al.}(2001)\textit{van Westrenen, Blundy, and
  Wood}}]{vWest:etal01a}
van Westrenen, W., J.~D. Blundy, and B.~J. Wood (2001), {High field strength
  element/rare earth element fractionation during partial melting in the
  presence of garnet: Implications for identification of mantle
  heterogeneities}, \textit{Geochem. Geophys. Geosyst.}, \textit{2}(7), 1039,
  \doi{10.1029/2000GC000133}.

\bibitem[{\textit{Wang and Takahashi}(2000)}]{WWaTa00}
Wang, W., and E.~Takahashi (2000), {Subsolidus and melting experiments of
  K-doped peridotite KLB-1 to 27\,GPa: Its geophysical and geochemical
  implications}, \textit{J. Geophys. Res.}, \textit{105}(B2), 2855--2868,
  \doi{10.1029/1999JB900366}.

\bibitem[{\textit{W{\"a}nke and Dreibus}(1994)}]{WaDr94}
W{\"a}nke, H., and G.~Dreibus (1994), {Chemistry and accretion history of
  Mars}, \textit{Phil. Trans. R. Soc. Lond.}, \textit{A 349}, 285--293.

\bibitem[{\textit{Watters et~al.}(2009)\textit{Watters, Zuber, and
  Hager}}]{WAWatt:etal09}
Watters, W.~A., M.~T. Zuber, and B.~H. Hager (2009), {Thermal perturbations
  caused by large impacts and consequences for mantle convection}, \textit{J.
  Geophys. Res.}, \textit{114}, E02001, \doi{10.1029/2007JE002964}.

\bibitem[{\textit{Werner}(2008)}]{Werner08}
Werner, S.~C. (2008), {The early martian evolution -- Constraints from basin
  formation ages}, \textit{Icarus}, \textit{195}(1), 45--60,
  \doi{10.1016/j.icarus.2007.12.008}.

\bibitem[{\textit{Werner and Ivanov}(2015)}]{WeIv15}
Werner, S.~C., and B.~A. Ivanov (2015), Exogenic dynamics, cratering, and
  surface ages, in \textit{Physics of Terrestrial Planets and Moons},
  \textit{Treatise on Geophysics}, vol.~10, edited by T.~Spohn, 2nd ed., chap.
  10.10, pp. 327--365, Elsevier, \doi{10.1016/B978-0-444-53802-4.00170-6}.

\bibitem[{\textit{Wieczorek}(2008)}]{Wieczorek08}
Wieczorek, M.~A. (2008), Constraints on the composition of the martian south
  polar cap from gravity and topography, \textit{Icarus}, \textit{196}(2),
  506--517, \doi{10.1016/j.icarus.2007.10.026}.

\bibitem[{\textit{Wieczorek and Zuber}(2004)}]{WiZu04}
Wieczorek, M.~A., and M.~T. Zuber (2004), {Thickness of the Martian crust:
  Improved constraints from geoid-to-topography ratios}, \textit{J. Geophys.
  Res.}, \textit{109}, E01009, \doi{10.1029/2003JE002153}.

\bibitem[{\textit{Williams and Nimmo}(2004)}]{JPWiNi04}
Williams, J.-P., and F.~Nimmo (2004), {Thermal evolution of the Martian core:
  Implications for an early dynamo}, \textit{Geology}, \textit{32}(2), 97--100.

\bibitem[{\textit{Wood and Blundy}(2002)}]{BJWoBl02}
Wood, B.~J., and J.~D. Blundy (2002), {The effect of H\textsubscript{2}O on
  crystal--melt partitioning of trace elements}, \textit{Geochim. Cosmochim.
  Acta}, \textit{66}(20), 3647--3656.

\bibitem[{\textit{Wood and Blundy}(2014)}]{BJWoBl14}
Wood, B.~J., and J.~D. Blundy (2014), Trace element partitioning: The
  influences of ionic radius, cation charge, pressure, and temperature, in
  \textit{The Mantle and Core}, \textit{Treatise on Geochemistry}, vol.~3,
  edited by R.~W. Carlson, 2nd ed., chap. 3.11, pp. 421--448, Elsevier,
  \doi{10.1016/B978-0-08-095975-7.00209-6}.

\bibitem[{\textit{Wood et~al.}(1999)\textit{Wood, Blundy, and
  Robinson}}]{BJWood:etal99}
Wood, B.~J., J.~D. Blundy, and J.~A.~C. Robinson (1999), {The role of
  clinopyroxene in generating U-series disequilibrium during mantle melting},
  \textit{Geochim. Cosmochim. Acta}, \textit{63}(10), 1613--1620.

\bibitem[{\textit{Zhao et~al.}(2009)\textit{Zhao, Zimmerman, and
  Kohlstedt}}]{YHZhao:etal09}
Zhao, Y., M.~E. Zimmerman, and D.~L. Kohlstedt, {Effect of iron content on the
  creep behavior of olivine: 1. Anhydrous conditions}, \textit{Earth Planet.
  Sci. Lett.}, \textit{287}(1--2), 229--240, \doi{10.1016/j.epsl.2009.08.006},
  2009.
  
\bibitem[{\textit{Zuber et~al.}(2000)\textit{Zuber, Solomon, Phillips, Smith,
  Tyler, Aharonson, Balmino, Banerdt, Head, Johnson, Lemoine, McGovern,
  Neumann, Rowlands, and Zhong}}]{Zube:etal00}
Zuber, M.~T., S.~C. Solomon, R.~J. Phillips, D.~E. Smith, G.~L. Tyler,
  O.~Aharonson, G.~Balmino, W.~B. Banerdt, J.~W. Head, C.~L. Johnson, F.~G.
  Lemoine, P.~J. McGovern, G.~A. Neumann, D.~D. Rowlands, and S.~Zhong (2000),
  {Internal structure and early thermal evolution of Mars from Mars Global
  Surveyor topography and gravity}, \textit{Science}, \textit{287}(5459),
  1788--1793, \doi{10.1126/science.287.5459.1788}.

\end{thebibliography}
\end{document}